\documentclass[fleqn,usenatbib]{mnras}

\usepackage{newtxtext,newtxmath}

\usepackage[T1]{fontenc}
\usepackage{ae,aecompl}
\usepackage{graphicx}
\usepackage{overpic}

\usepackage{graphicx}	
\usepackage{amsmath}	
\usepackage{amssymb}	
\usepackage[usenames, dvipsnames]{color}

\newcommand{\kms}{\,{\rm km\,s^{-1}}}
\newcommand{\msun}{\,{\rm M_\odot}}

\newcommand{\beq}{\begin{equation}}
\newcommand{\eeq}{\end{equation}}
\newcommand{\ba}{\begin{eqnarray}}
\newcommand{\ea}{\end{eqnarray}}
\newcommand{\all}{\emph{all}}
\newcommand{\dx}{\emph{8dx}}
\newcommand{\rsm}{\emph{05Reff}}
\newcommand{\rbg}{\emph{sel}}

\newcommand{\hagn}{\mbox{{\sc \small Horizon-AGN}}}
\newcommand{\nh}{\mbox{{\sc \small NewHorizon}}}

\title[MBH mergers across mass scales]
{Black hole mergers from dwarf to massive galaxies with the NewHorizon and Horizon-AGN simulations}

\author[M. Volonteri et al.]{Marta Volonteri,$^{1}$\thanks{E-mail: martav@iap.fr (MV)}
Hugo Pfister,$^{2,3}$
Ricarda S. Beckmann,$^{1}$
Yohan Dubois,$^{1}$ \newauthor Monica Colpi,$^{4,5}$ Christopher J. Conselice,$^6$ Massimo Dotti,$^{4,5}$ Garreth Martin,$^{7,8}$ \newauthor
Ryan Jackson,$^{9}$ Katarina Kraljic,$^{10}$ Christophe Pichon,$^{1,11}$ Maxime Trebitsch,$^{12,13}$ \newauthor 
Sukyoung. K. Yi,$^{14}$  Julien Devriendt,$^{15}$ S\'ebastien Peirani$^{16,1}$\\
$^{1}$Institut d'Astrophysique de Paris, Sorbonne Universit\'e, CNRS, UMR 7095, 98 bis bd Arago, 75014 Paris, France\\
$^{2}$DARK, Niels Bohr Institute, University of Copenhagen, Blegdamsvej 17, DK-2100 Copenhagen, Denmark\\
$^{3}$Department of Physics, The University of Hong Kong, Pokfulam Road, Hong Kong, China\\
$^4$Dipartimento di Fisica ``G. Occhialini'', Universit\`a degli Studi di Milano-Bicocca, Piazza della Scienza 3, I-20126 Milano, Italy\\
$^5$INFN, Sezione di Milano-Bicocca, Piazza della Scienza 3, I-20126 Milano, Italy\\
$^6$Centre for Astronomy and Particle Theory, University of Nottingham, University Park, Nottingham, NG7 2RD, UK\\
$^{7}$Steward Observatory, University of Arizona, 933 N. Cherry Ave, Tucson, AZ, USA\\
$^{8}$Korea Astronomy and Space Science Institute, 776 Daedeokdae-ro, Yuseong-gu, Daejeon 34055, Korea.\\
$^{9}$Centre for Astrophysics Research, School of Physics, Astronomy and Mathematics, University of Hertfordshire, Hatfield, AL10 9AB, UK\\
$^{10}$Institute for Astronomy, Royal Observatory, Edinburgh EH9 3HJ, United Kingdom\\
$^{11}$Korea Institute of Advanced Studies (KIAS) 85 Hoegiro, Dongdaemun-gu, Seoul, 02455, Republic of Korea\\
$^{12}$Max-Planck-Institut f{\"u}r Astronomie, K{\"o}nigstuhl 17, 69117 Heidelberg, Germany\\
$^{13}$Zentrum f{\"u}r Astronomie der Universit{\"a}t Heidelberg, Institut f{\"u}r Theoretische Astrophysik, Albert-Ueberle-Str. 2, 69120 Heidelberg, Germany\\
$^{14}$Department of Astronomy and Yonsei University Observatory, Yonsei University, Seoul 03722, Republic of Korea\\
$^{15}$University of Oxford, Astrophysics, Denys Wilkinson Building, Oxford OX1 3RH, UK\\
$^{16}$Universit\'e C\^ote d'Azur, Observatoire de la C\^ote d'Azur, CNRS, Laboratoire Lagrange, Nice, France}

\date{Accepted XXX. Received YYY; in original form ZZZ}

\pubyear{2020}

\begin{document}
\label{firstpage}
\pagerange{\pageref{firstpage}--\pageref{lastpage}}
\maketitle

\begin{abstract}

Massive black hole (MBH) coalescences are powerful sources of low-frequency gravitational waves. To study these events in the cosmological context we need to trace the large-scale structure and cosmic evolution of a statistical population of galaxies, from dim dwarfs to bright galaxies. To cover such a large range of galaxy masses,  we analyse two complementary simulations: \hagn~with a large volume and low resolution which tracks the high-mass ($>10^7\msun$) MBH population, and \nh~with a smaller volume but higher resolution that traces the low-mass ($<10^7\msun$) MBH population. While \hagn~can be used to estimate the rate of inspirals for Pulsar Timing Arrays, \nh~can investigate MBH mergers in a statistical sample of dwarf galaxies for LISA, which is sensitive to low-mass MBHs.
  We use the same method to analyse the two simulations, post-processing MBH dynamics to account for time delays mostly determined by dynamical friction and stellar hardening.  In both simulations, MBHs typically merge long after galaxies do, so that the galaxy morphology at the time of the MBH merger is no longer determined by the  structural disturbances engendered by the galaxy merger from which the MBH coalescence has  originated. These time delays cause a loss of high-$z$ MBH coalescences, shifting the peak of the MBH merger rate to $z\sim 1-2$. 
  This study shows how tracking MBH mergers in low-mass galaxies is crucial to probing the MBH merger rate for LISA and investigate the properties of the host galaxies. 
\end{abstract}




\section{Introduction}
Black holes are astrophysical objects that emit radiation over the whole electromagnetic spectrum, from the radio to gamma-rays. When their spacetime is highly dynamical, such as in mergers with other black holes, they also emit gravitational waves, as spectacularly demonstrated following the ground-braking discovery of the first stellar black hole binary by the LIGO and Virgo collaborations \citep{2016PhRvL.116f1102A}. 

The frequency at which black holes emit gravitational waves depends on the inverse of the binary mass. To a first approximation the frequency at coalescence is close to the Keplerian frequency of a test particle revolving at the innermost stable circular orbit around a black hole with mass equal to the mass of the binary. While LIGO-Virgo cannot detect gravitational waves from mergers of black holes much more massive than about $100 \msun$ \citep{Mangiagli2019}, experiments with a much longer baseline can detect them.  The ESA space mission LISA \citep{2017arXiv170200786A} aims at detecting low-frequency, $\sim 10^{-4}-1$ mHz, gravitational waves from the coalescence of massive black holes (MBHs) with masses $\sim 10^4-10^7 \msun$,  the TianQin project aims detection at a similar but somewhat reduced mass range \citep{Luo_2016,PhysRevD.99.123002,2019PhRvD.100d4036S}, while Pulsar Timing Array experiments \citep[PTA,][]{2004ApJ...606..799J,2005ApJ...625L.123J} at nHz frequency  probe the monochromatic inspiral of tight MBH binaries with mass $>10^8 \msun$. LISA- and PTA-detectable MBHs are those black holes that inhabit the centres of massive galaxies and that participated in cosmic evolution by shining as quasars. 

Gravitational wave experiments alone, as well as multi-messenger studies including electromagnetic counterparts, open the possibility of obtaining complementary and unique information on the evolution of MBHs. These studies will allow us to discover MBH ``seeds'' at high-redshift \citep{GW3}, to obtain information on the dynamical evolution of massive black holes \citep{2019MNRAS.486.4044B}, to assess the relative role of accretion and mergers in establishing scaling relations between MBHs and galaxies \citep{VN09}, to modify MBH spins \citep{BertiVolonteri2008} and to gain information on the properties of gas in an accretion disc \citep{2019MNRAS.486.2754D}. 

Predictions and estimates of the merger rate and the properties of merging MBHs have developed over the years, starting with analytical models \citep{Haehnelt1994,Jaffe2003,Wyithegw03}, then with semi-analytical models \citep{Sesanaetal2004,Tanaka2009,Sesana2011,2012MNRAS.423.2533B,2018MNRAS.481.3278R}, and more recently with hydrodynamical cosmological simulations \citep{2016MNRAS.463..870S,2017MNRAS.464.3131K,2020MNRAS.491.2301K}. 
Analytical models can only study a population of MBHs without information on single sources. Semi-analytical models improve on that, but they lack spatial information and use simplified analytical functions. However, both approaches have high flexibility and small computational costs, allowing for parametric studies. Vice-versa, cosmological hydrodynamical simulations naturally include spatial information of galaxies and can reach a high-level of complexity, but at high computational costs. 

The landscape of cosmological hydrodynamical simulations currently comprises low-resolution large-volume simulations with many massive galaxies, but which are unable to resolve dwarf galaxies, or high-resolution small-volume simulations with few galaxies, but which are capable of probing the dwarf regime.  The former simulations are well adapted to studying the high-mass end of the merging MBH population, $10^7-10^8 \msun$ and higher, i.e. the binaries that are relevant for Pulsar Timing Array experiments  \citep{PTA2009,2017MNRAS.464.3131K}, but they cannot be used to study the full LISA MBH population, $10^4-10^7 \msun$, since such MBHs are hosted in galaxies unresolved in these simulations. The latter type of simulations can resolve the merging history of the galaxies hosting LISA MBHs, but at the expense of smaller statistical significance. 

More specifically, most large-scale hydrodynamical cosmological simulations have relatively low mass resolution, typically resolving galaxies with mass $>10^9\msun $ ~\citep{2014MNRAS.444.1453D,Vogelsberger2014,2015MNRAS.446..521S,Pillepich2018,Dave2019}. They cannot resolve galaxies where low-mass MBHs are expected to reside, or they implant MBH seeds only in high-mass haloes, thus missing the early merger history \citep{2014MNRAS.445..175G,2015MNRAS.446..521S,2016MNRAS.458.1013S}. The low spatial resolution of $\sim 1$~kpc makes it challenging to account for the MBH dynamics. Simulations with intermediate volumes and mass resolutions have started to uncover the population of merging MBHs in lower-mass haloes \citep{2018MNRAS.475.4967T}.  Alternatively, small-scale zooms can resolve the evolution of galaxy pairs in small groups with high spatial resolution \citep{2016ApJ...828...73K}, but do not provide statistical samples.

In this paper we analyse two simulations: one, Horizon-AGN~\citep[\hagn,][]{2014MNRAS.444.1453D} is a low-resolution large-volume cosmological simulation similar to that used for previous studies by other groups \citep{2016MNRAS.463..870S,2020MNRAS.491.2301K}. This simulation is well-adapted for studying the high-mass end of the MBH population residing in massive galaxies, and can be used to investigate MBHs in the mass and redshift of interest for PTAs. The second simulation, \nh~(Dubois et al. in prep.) is a high-resolution zoom that can resolve dwarf galaxies with 40 pc resolution so that bulges and discs can be physically resolved~\citep{2019ApJ...883...25P}, in a sufficiently large volume such that the simulation includes more than one hundred galaxies with stellar mass larger than $10^9\msun$ at the final redshift analysed here, $z=0.45$. \nh~is well suited for studying LISA's MBHs. \nh~is the first cosmological simulation of this kind used for studying the MBH merger rate in low-mass galaxies.

Even with the extremely high resolution of \nh, corrections to the merger timescales are necessary because in reality MBHs merge when their separation is comparable to their gravitational radius, $10^{-6} {\rm pc} (M_{\rm BH}/10^7 \msun)$,  and cosmological simulations are unable to obtain such resolution. General-relativistic simulations of black hole mergers that resolve the full spacetime, conversely, are limited to evolving only a few orbits and have a computational domain of only a few hundreds of gravitational radii \citep[see][for a compendium]{2014CQGra..31k5004A}.  

We analyse MBH mergers in the two simulations in the same way, explore the differences and similarities and discuss their relative advantages and disadvantages. We also connect MBH mergers to galaxy mergers, both one by one and in a statistical sense, comparing the merger rates, and the masses and mass ratios of galaxy mergers sourcing MBH mergers to the general merging population in order to assess how to infer information on MBH mergers from galaxy merger samples.

\section{Horizon-AGN and NewHorizon}
\label{sec:simulations}

Both simulations analysed in this paper form part of the HORIZON simulation suite\footnote{https://www.horizon-simulation.org/}. \hagn~\citep{2014MNRAS.444.1453D} is a hydrodynamical cosmological simulation with a box length of $142 \rm \, Mpc$, and \nh~(NH\footnote{https://new.horizon-simulation.org/}, Dubois et al. in prep) is a zoom-in simulation that resimulated a sub-sphere with a radius of 10 comoving Mpc of \hagn~at higher resolution. 

The properties of the MBH population in \hagn~is in good agreement with observations of  MBHs with masses $M_{\rm BH}>10^7 \msun$ \citep{2016MNRAS.460.2979V}, and therefore this is the type of simulation that can be used to probe the MBH mergers of interest for PTA.  The most massive MBHs in \nh~have mass $\sim 10^7 \msun$ instead, therefore this simulation probes the MBHs of interest for LISA. A full analysis of the MBH population in \nh~will be presented in Beckmann et al. (in prep), but we note that it produces an AGN luminosity function in good agreement with the faint end of the observed one at $z=0.5-3$. 

Both simulations are based on a standard $\Lambda$CDM cosmology consistent with WMAP-7 data \citep{Komatsu11}, with total matter density $\Omega_{\rm m}=0.272$, dark energy density $\Omega_{\Lambda}=0.728$, baryon density $ \Omega_{\rm b}=0.045$, a Hubble constant of $H_{0}=70.4 \rm \ km\,s^{-1}\, Mpc^{-1}$, and an amplitude of the matter power spectrum and power-law index of the primordial power spectrum of $\sigma_8=0.81$ and $n_{\rm s}=0.967$ respectively.
The two simulations rely on the same sampling of the initial phases (white noise) generated at the common spatial resolution $4096^3$ (\nh) and then downgraded at lower resolution $1024^3$ (\hagn).

The simulations were run with the adaptive mesh refinement code {\sc ramses} \citep{teyssier02}, using a second-order unsplit Godunov scheme to solve the Euler equations, and an HLLC Riemann solver with a MinMod Total Variation Diminishing scheme to reconstruct interpolated variables. Both simulations have adaptive mesh refinement performed in a quasi-Lagrangian manner if the total mass in a cell becomes greater than 8 times the initial mass resolution. In \nh~we have also added a Jeans refinement criterion to refine the mesh if a cell has a length shorter than one Jeans length where the gas number density is larger than $5\, \rm H\, cm^{-3}$. Collisionless particles (dark matter and star particles) are evolved using a particle-mesh solver with a cloud-in-cell interpolation. 

The two simulations have been run with different sub-grid physics, and therefore they cannot be combined to have a ``full'' merger rate: when discussing results we refrain from merging the results of the simulations, and we keep them separate.  We summarise below the main characteristics of the implementations in the two simulations and we refer to  \cite{2014MNRAS.444.1453D} and \cite{2019ApJ...883...25P} for more details. 

\subsection{Horizon-AGN}
\label{sec:HAGN}

\hagn~simulates a large volume, (142 comoving Mpc)$^3$, at a relatively low spatial and mass resolution: refinement is permitted down to $\Delta x=1$~kpc, the dark matter particle mass is $8\times 10^7 \msun$, the stellar particle mass is $2\times 10^6 \msun$, and the MBH seed mass is $10^5 \msun$. 

Gas cooling is modelled down to $10^4\, \rm K$ with curves from \cite{sutherland&dopita93}. Heating from a uniform UV background takes place after redshift $z_{\rm reion} = 10$ following \cite{haardt&madau96}. The gas follows an equation of state for an ideal monoatomic gas with an adiabatic index of $\gamma_{\rm ad}=5/3$.

Star formation is modelled with a Schmidt relation adopting a constant star formation efficiency $\epsilon_*=0.02$ \citep{kennicutt98, krumholz&tan07} in regions which exceed a gas hydrogen number density threshold of $n_0=0.1\, \rm H\, cm^{-3}$ following a Poisson random process \citep{rasera&teyssier06, dubois&teyssier08winds}. Mechanical energy injection from Type Ia SNe, Type II SNe and stellar winds is included assuming a \citet{1955ApJ...121..161S} initial mass function with cutoffs at $0.1\, \rm M_{\odot}$  and $100 \, \rm M_{\odot}$. 

MBHs are created in cells where the gas density is larger than $n_0$, and where the gas velocity dispersion is larger than $100\, \rm km\,s^{-1}$; an exclusion radius of 50 comoving kpc is imposed to avoid formation of multiple MBHs in the same galaxy. MBH formation is stopped at $z=1.5$.  The accretion rate follows a Bondi-Hoyle-Littleton rate modified by a factor $\alpha=(n/n_0)^2$ when $n>n_0$ and $\alpha=1$ otherwise \citep{Booth2009} in order to account for the inability to capture the multiphase nature of the interstellar gas. The effective accretion rate onto MBHs is capped at the Eddington luminosity with a radiative efficiency of 0.1.   For luminosities above 1\% of the Eddington luminosity, 15\% of the MBH emitted luminosity is isotropically coupled to the gas within $4\Delta x$ as thermal energy. 

At lower luminosities feedback takes a mechanical form, with 100\% of the power injected into a bipolar outflow with a velocity of $10^4\,\rm km\, s^{-1}$, injected in a cylinder with radius $\Delta x$ and height $2 \, \Delta x$. 

To avoid spurious motions of MBHs due to finite force resolution effects, we adopt an explicit drag force of the gas onto the MBH \citep{2012MNRAS.420.2662D} with the same boost factor $\alpha$ used for accretion.   This gas dynamical friction is expressed as $F_{\rm DF}= f_{\rm gas} 4 \pi \alpha \rho_{\rm gas} (G M_{\rm BH}/\bar c_s)^2$, where $\rho_{\rm gas}$ is the mass-weighted mean gas density within a sphere of radius $4 \, \Delta x$ and $f_{\rm gas}$ is a factor function of the mach number ${\mathcal M}=\bar u/\bar c_s$, which accounts for the extension and shape of the wake \citep{Ostriker1999} and takes a value between 0 and 2 for an assumed Coulomb logarithm of 3 \citep{chaponetal13}. Including this drag force allows us to avoid pinning MBHs to galaxy centers, i.e.  constantly repositioning them at the minimum of the local potential, which causes unnatural dynamics \citep{2015MNRAS.451.1868T}. See \cite{Dubois2013} for additional details.

\subsection{NewHorizon}
\label{sec:NH}
\nh~is a smaller and higher resolution volume: it resimulates an ``average'' density sphere with a radius 10 comoving Mpc of Horizon-AGN, focusing on field galaxies. The most massive halo at $z=0.45$ is $5.9\times10^{12} \msun$. The dark matter mass resolution in the zoomed region is $M_{\rm DM,hr}=1.2\times 10^6 \,\rm M_\odot$, stellar mass resolution is $10^4 \,\rm M_\odot$, $\Delta x=40$~pc, and the MBH seed mass $10^4 \msun$.  In the high-resolution region a passive refinement scalar is injected, with a value of $1$ within that region, and we only allow for refinement when the value of this scalar is above\footnote{The initial pure Lagrangian volume changes size and shape over time and the Eulerian volume gets polluted by low-resolution dark matter particles over time.} a value of $0.01$. We use this passive scalar also to measure the refined volume, which collapses somewhat from the initial volume of $\sim 4000$ Mpc$^3$ under the effect of gravity.

Gas cooling is modelled with rates tabulated by \cite{sutherland&dopita93} above $10^4\,\rm K$ and those from \cite{dalgarno&mccray72} below $10^4\,\rm K$. The uniform UV background is as in Horizon-AGN, except that we include self-shielding where the gas density is larger than $n_{\rm shield}=0.01\,\rm H\, cm^{-3}$, i.e. UV photo-heating is reduced by a factor $\exp(-n/n_{\rm shield})$.

Star formation follows a Schmidt law, but with a density threshold of $n_0=10\, \rm H\, cm^{-3}$, and with a varying star formation efficiency that depends on the local turbulent Mach number and Jeans length~\citep[see e.g.][]{Trebitsch2018}. The initial mass function differs from Horizon-AGN, adopting a Chabrier functional form \citep{2005ASSL..327...41C}, with cutoffs at 0.1 and 150~$\msun$. We include the mechanical SN feedback model from \cite{kimmetal15}, which models the energy and momentum conserving phases of the explosion separately. This SN feedback is more effective than the purely kinetic formulation used in Horizon-AGN~\citep{kimmetal15}.

Black holes form in cells where both the gas and stellar densities are above the threshold for star formation and the stellar velocity dispersion is larger than $20\,\rm km\,s^{-1}$, again with an exclusion radius of 50 comoving kpc. An explicit drag force of the gas onto MBHs is implemented, with the same boost as in Horizon-AGN. Since \nh~has sufficient spatial resolution to model some of the multiphase nature of the gas, accretion is modelled via a non-boosted Bondi-Hoyle-Littleton rate ($\alpha=1$), capped at the Eddington luminosity. 

Spin evolution via gas accretion and MBH-MBH mergers are followed explicitly in \nh, using the implementation described in \cite{2014MNRAS.440.2333D}. For MBH-MBH mergers we adopt for the final spin a fit motivated by general-relativistic simulations \citep{rezzollaetal08}. For accretion, the direction of the angular momentum of the accreted gas is used to decide whether the accreted gas feeds an aligned or misaligned disc \citep{King2005}, respectively spinning up or down the MBH \citep{Bardeen1970} for accretion above 1\% of the Eddington luminosity. At lower luminosities the MBH spin up (down) rate is motivated by jets tapping energy from MBH spins \citep{BZ1977,Moderski1996} and follows the results from \cite{mckinneyetal12}, where we fit  a fourth-order polynomial to their sampled values (from their table 7, A$a$N100 runs, where $a$ is the value of the MBH spin). The radiative efficiency is calculated individually for each MBH based on its spin, thus the Eddington mass accretion rate will vary as a MBH spin evolves, and it is reduced linearly with the Eddington ratio for MBHs below 1\% of the Eddington luminosity following \cite{benson&babul09}. AGN feedback also depends on spin, being 15\% of the spin-dependent emitted power for the thermal feedback, injected within a sphere of radius $\Delta x$, and following the results of magnetically chocked accretion discs of  \citet[][from their table 5 for A$a$N100 runs]{mckinneyetal12}. The bi-polar outflows are modelled as in Horizon-AGN. We also enforce the refinement within a region of radius $4\Delta x$ around the MBH at the maximum allowed level of refinement.

\subsection{Halo and galaxy catalogues}
In \hagn~we identify dark matter haloes and galaxies with the AdaptaHOP halo finder \citep{aubertetal04}. The density field used in AdaptaHOP is smoothed over 20 particles. In \hagn~we fix the density threshold at 178 times the average total matter density and require 50  dark matter particles for identification for both dark matter haloes and 50 star particles for galaxies. In \nh~only haloes with average density larger than 80 times the critical density are considered, and the minimum number of particles per halo is 100.
For galaxies in \nh, we select them using the HOP finder, requiring at least 50 star particles for a galaxy. 
The difference with AdaptaHOP is that substructures are not extracted from the main component. 
This is necessary as the high resolution achieved in the \nh~galaxies allows the formation of dense star forming clumps, which will otherwise be removed with AdaptaHOP. 

 \nh~is a zoom simulation embedded in a larger cosmological volume filled with lower resolution dark matter particles, we need to consider only ``pure'' haloes, devoid of low resolution dark matter particles, and the galaxies embedded in these haloes.  Low-resolution particles are more massive than high-resolution particles, therefore the presence of low- and high-resolution particles in the same halo would induce numerical issues; using only pure galaxies/halos the dynamics of the galaxies under examination is robust. The purity criterion only affects halos at the edge of the refined volume, where lower-resolution particles from outside can get mixed into the high-resolution region. At the final redshift of  $z=0.45$, this gives 762 uncontaminated galaxies with stellar mass larger than $5\times 10^5\,\rm M_\odot$ and 238, 87 and 17 with masses larger than $10^8 \msun$, $10^9\msun$ and $10^{10}\msun$ respectively. 

In Horizon-AGN, we obtain the centre of haloes and galaxies using the shrinking sphere approach proposed by \cite{poweretal03} to get the correct halo centre. In \nh~we use the shrinking sphere approach for haloes and for galaxies with a reduction factor of the radius of respectively 10 and 30 per cent at each shrinking step, and with a stopping search radius of respectively 0.5 and 1 kpc. This rather large value for galaxies (with respect to their typical size) in \nh~allows us to avoid converging towards dense and massive but randomly located clumps, and provides a qualitatively better centring over the total stellar distribution. Despite this, defining the galaxy ``centre'' for low-mass and high-z galaxies remains challenging.

\section{Methods}
\label{sec:methods} 

\subsection{Selecting black hole mergers}
\label{sec:selection} 

 Using the information in the simulation and post-processing techniques we create catalogues of MBH mergers. To select merging MBHs, we use the information on all MBHs at each synchronised (coarse) timestep of the corresponding simulation, every $\sim 0.5$ Myr in \nh~and $\sim 0.6-0.7$ Myr in Horizon-AGN. Only MBH information is available at each coarse timestep, as galaxy and halo information is only available at full outputs which are saved much less frequently.

A pair of MBHs are merged \emph{in the simulation} when a MBH present at a given time disappears in the succeeding time step; the companion MBH is located by searching for the MBH that was closest to the vanished one, keeping in mind that the simulation merges MBHs when they are separated by $\leqslant 4 \Delta x$, corresponding to 4~kpc for \hagn~and 160~pc for \nh, and they are energetically bound in vacuum. These separations, especially for \hagn, are much larger than those where MBHs effectively merge in normal conditions and in section~\ref{sec:delays} we discuss how we include additional timescales to account for this.  We refer to these as ``numerical mergers'' and the initial lists contains 542 MBH mergers for \nh~and 85397 for Horizon-AGN. After a numerical merger, the remnant acquires the ID of the most massive MBH in the pair. This is the starting general catalogue for \hagn~(``\all'') and we use the subscript ``${\rm in}$'' for quantities (redshift, cosmic time, MBH masses, accretion rates, galaxy masses) at the initial time of the numerical merger. 

For \nh, which is a zoom simulation, this initial list also includes MBHs that are located in polluted haloes, i.e., haloes that contain low-resolution dark matter particles and whose evolution is therefore untrustworthy. The first task is therefore to remove MBHs in polluted haloes from our sample. This is performed by spatially matching MBHs with galaxies in unpolluted haloes. For each MBH pair we request that the primary is located within $\max(10R_{\rm eff},4\Delta x)$ from a galaxy in an unpolluted halo, where $R_{\rm eff}$ is the geometric mean of the projected half-mass radius over the 3 cartesian axes. A galaxy is associated to a halo if the galaxy centre is within 10\% of the virial radius of the dark matter halo. The initial list of 542 numerical mergers is thus reduced to 385. For \nh~we consider this to be the starting general catalogue (``\all'').   

In both \nh~and Horizon-AGN, MBHs are not pinned in the centre of galaxies and haloes, so an additional concern is removing possible ``spurious'' numerical mergers. In a realistic situation it is unlikely that MBHs merge far from the centre of the host galaxy \citep{Merritt2001}. This is because MBHs have a small impact parameter for binding to another MBH: they must pass within each other's sphere of influence.  In the simulation, for slowly moving MBHs the impact parameter is instead $4 \Delta x$, much larger than the physical impact parameter under realistic conditions. The energy condition ensures that in vacuum the MBHs would bind, but not that they would merge, since much energy has to be extracted from the binary before they are sufficiently close that gravitational waves become effective in driving the MBHs to coalescence. Processes that extract energy from the orbit of a MBH or from a binary, first dynamical friction, then hardening by scattering off single stars and torques from  circumbinary disc, are far more effective in higher surrounding stellar and/or gas densities. For off-centre MBHs and binaries, the orbital evolution is therefore slower than for centrally located MBHs. This is particularly important for \nh, where the low initial MBH seed mass can make dynamics erratic \citep{2019MNRAS.486..101P}, while for \hagn~the main worry, as we will see below, is the correction for dynamical delays, even in the case of ``central" mergers. 

To identify possible spurious numerical mergers we cross-correlate MBHs with galaxies, and we select pairs where the MBHs are within $\max(2R_{\rm eff},4\Delta x)$ (``\rbg'').  In \nh~galaxy properties are output every $\sim 15$ Myr, while in \hagn~every $\sim 150$ Myr. In \nh~we consider the galaxy output closest in time to the numerical merger. Given the sparsity of \hagn~outputs, galaxies and MBHs can have moved considerably between the two outputs surrounding the time of the MBH merger; for this simulation we apply the criterion $\max(2R_{\rm eff},4\Delta x)$ on the primary MBH at its position in the galaxy outputs before and after the merger.

In \nh, this catalogue contains 314 MBHs, and in \hagn~it contains  64505 MBHs. Numerical noise in MBH dynamics arises when the ratio of MBH mass to stellar particle mass is too small, which results in a lack of dynamical friction from particles \citep{2015MNRAS.451.1868T,2019MNRAS.486..101P} that would otherwise offset this noise.  This spurious noise can keep MBHs away from galaxy centres, and, hence, we define ``centre'' very generously and we consider \rbg~as the reference sample. We discuss alternative choices in  Appendix~\ref{app:spurious}, noting that choices can make a significant difference -- up to an order of magnitude -- in the final results.  The merger rate is still likely to be considered a lower limit. We apply the same selection criterion of  $\max(2R_{\rm eff},4\Delta x)$ in matching MBHs to galaxies at all subsequent steps of the post-processing described below.

\subsection{Dynamical evolution}
\label{sec:delays} 

As discussed, the simulations numerically ``merge'' MBHs when their separation is much larger than the typical separation, of order of a millipc, where emission of gravitational waves becomes effective and brings the binary to coalescence in less than a Hubble time \citep{Colpi2014}. The catalogues have to be post-processed to obtain more realistic merging timescales and merger rates.  Various approaches for adding merger timescales in semi-analytical models \citep{VHM,2012MNRAS.423.2533B,2019MNRAS.486.4044B} and in numerical simulations  by post-processing have been proposed \citep{2017MNRAS.464.3131K,2019ApJ...879..110K,2020MNRAS.491.2301K,2020arXiv200606647S}. Any correction will necessarily be more arbitrary for low-resolution simulations, since the available information must be extrapolated over a much larger parameter space.

\subsubsection{Dynamical friction}
 The simulation includes a drag force from surrounding gas calculated on the fly, but it only operates above the resolution limit. For numerically merged MBHs the drag would act on the ``merged'' MBH and not on each MBH in the pair individually. This part of the dynamical decay requires additional dynamical friction in post-processing.
A typical approach is to include the dynamical friction timescale for the secondary MBH to infall to the galaxy centre, where the primary MBH is assumed to sit at rest.
In \nh, most MBHs are not growing much until they reside in galaxies with typical stellar mass above $10^{10}\,\rm \msun$ due to efficient feedback from SNe removing gas from the innermost regions of galaxies~\citep{2015MNRAS.452.1502D,Habouzit2017,Bower2017}.
Therefore, for \nh, since many MBHs are not growing much, very often the two merging MBHs have very similar masses and there is no clear distinction between  ``primary'' and ``secondary''.  We can therefore imagine that both MBHs have to find the centre of the galaxy.

We here take a relatively simple approach. We first estimate dynamical friction evolution as in \citet{2019ApJ...879..110K} by computing  the frictional timescale for a massive object in an isothermal sphere \citep{binney1987}: 
\begin{equation}
t_{\rm df}=0.67\, {\rm Gyr} \left(\frac{a}{4\, {\rm kpc}}\right)^2\left(\frac{\sigma}{100 \, {\rm km\, s^{-1}}}\right)\left(\frac{M_{\rm BH}}{10^8 \,M_{\odot}}\right)^{-1}\frac{1}{\Lambda},
\label{eq:tdf}
\end{equation}
where $M_{\rm BH}$  is the black hole mass, $\sigma$ the central stellar velocity dispersion approximated as $(0.25GM_{\rm gal}/R_{\rm eff})^{1/2}$ and $\Lambda=\ln(1+M_{\rm gal}/M_{\rm BH})$, with $M_{\rm gal}$  the total stellar mass of the galaxy hosting the MBH at the output closest in time.  In Eq.~\ref{eq:tdf}, $a$ is the distance of the MBH from the galaxy centre, calculated  in \nh\ at the output closest in time to the numerical merger.  In \hagn, where galaxy outputs are much sparser, we first interpolate linearly the position of the host galaxy  within the box from the output before the merger and the output after the merger to the time of the numerical merger and then calculate the distance between each of the MBHs and the galaxy center at that time. In the normalization of $t_{\rm df}$ we included a factor 0.3 to account for typical orbits being non-circular \citep{Taffoni2003}. We calculate the sinking time  for $M_1$, the most massive MBH in the pair, and $M_2$, the least massive, and take the longest of the two, which is normally associated to $M_2$ (in \hagn~as in \nh, the masses are usually similar except in the most massive galaxies). 
The difficulty in matching MBHs to galaxies due to the discrete temporal sampling of the galaxy outputs induces scatter in the estimates of the this timescale, which can be either under- or over-estimated. 

We have not included additional corrections for stellar mass bound to the MBH, which would speed up the orbital decay \citep{Taffoni2003,Callegari2009,Svanwas2012}, nor corrections for the increase in mass of the MBHs due to accretion and for the increase in the stellar mass and velocity dispersion of the galaxy, during this time.  In Appendix~\ref{app:evolution}  we present an analysis of the impact of our assumptions on the estimate of the dynamical friction timescale, for the last two effects mentioned above. We briefly note here that black hole growth does not seem significant and that the potential impact of a rising velocity dispersion is of decreasing the dynamical friction timescale.

\subsubsection{Binary evolution}

After the dynamical friction timescale has elapsed, we look in the MBH outputs for the ID of the remnant MBH at $t_{\rm in}+t_{\rm df}$, where $t_{\rm in}$ is the time of the numerical merger.  If at $t_{\rm in}+t_{\rm df}$   the MBH descendant of the numerical merger is within $\max(2R_{\rm eff},4\Delta x)$ of a galaxy, we calculate additional delay timescales. If $t_{\rm df}$ is very long, typically for low-mass MBHs at large distances from galaxy centres, the MBH may have already experienced another numerical merger as secondary MBH.  We take the merging product between the original MBH and the new primary MBH and proceed as in the previous case. 

 We assume that after $t_{\rm df}$ has elapsed, the MBHs form a close binary with total mass $M_{12}$ equal to the mass of the MBH in the simulation at $t_{\rm in}+t_{\rm df}$, thus including the mass accreted during the dynamical friction phase, and and mass ratio $q$ equal to that of $M_2/M_1$ at the time of the numerical merger. We consider the MBH and host galaxy properties at $t_{\rm in}+t_{\rm df}$ to calculate ensuing binary evolution timescales. The main dynamical drivers are either hardening by individual scattering off stars or viscous interactions in a circumbinary disc\footnote{Various groups find that torques in circumbinary discs cause the binary to outspiral rather than inspiral, especially for binary mass ratios close to unity \citep{2017MNRAS.466.1170M,2019ApJ...875...66M,2019ApJ...871...84M,2019arXiv191105506D}. We do not take this into account, but we  note that in most cases the dynamical evolution of the binaries in our simulation is dominated by stellar scattering for sufficiently dense stellar structures.}.

We calculate the two timescales following \cite{2015MNRAS.454L..66S} for the stellar hardening and \cite{2015MNRAS.448.3603D} for gaseous torques between the MBH binary and the circumbinary disc.  We calculate the former as
\beq
t_{\rm bin,h}=15.18\, {\rm Gyr} \left(\frac{\sigma_{\rm inf}}{\kms}\right) \left(\frac{\rho_{\rm inf}}{\msun {\rm pc}^{-3}}\right)^{-1}\left(\frac{a_{\rm gw}}{10^{-3}\rm{pc}}\right)^{-1}, 
\label{eq:tbinh}
\eeq
where $\sigma_{\rm inf}$ and $\rho_{\rm inf}$ are the velocity dispersion and stellar density at the sphere of influence, defined as the sphere containing twice the binary mass in stars, and $a_{\rm gw}$ is the separation at which the binary spends most time, i.e., where the effectiveness of hardening and gravitational wave emission is the lowest:

\begin{equation}
\begin{split}
a_{\rm gw} = & 2.64 \times 10^{-2} {\rm pc} \, \times \\
 & \left[\frac{\sigma_{\rm inf}}{\kms}\,\frac{\msun {\rm pc}^{-3}}{\rho_{\rm inf}}\,\frac{15}{H}\,\left (\frac{M_1\,M_2\,M_{12}}{2\times10^{24} \rm M^3_\odot}\right )\right]^{1/5},
\end{split}
\label{eq:agw}
\end{equation}
 where $M_{12}=M_1+M_2=M_{\rm BH}$ is the mass of the binary and we adopt $H=15$ as a reference value  \citep{2015MNRAS.454L..66S}, although in some environments $H$ can be non-constant \citep{2019arXiv191111526O}. 
 
 To estimate $\sigma_{\rm inf}$ and $\rho_{\rm inf}$ we assume that the density profile within $R_{\rm eff}$ is a power-law $\rho \propto r^{-\gamma}$ with index $\gamma=2$ (See Appendix~\ref{app:gamma} for a comparison with $\gamma=1$), with a total mass within $R_{\rm eff}$ equal to $1/2$ the galaxy stellar mass. Given the power-law profile and $M_{\rm BH}$ one can calculate the radius $r_{\rm inf}$ containing twice the binary mass in stars, the density at that radius, $\rho_{\rm inf}$, and $\sigma_{\rm inf}$ as $(GM_{\rm BH}/r_{\rm inf})^{1/2}$. Neither simulation can resolve nuclear star clusters, in the presence of which $t_{\rm bin,h}$ becomes much shorter \citep{2018MNRAS.477.4423A,2019MNRAS.487.4985B,2019arXiv191111526O}, therefore for galaxies with mass $\sim10^8-10^{10} \msun$ where the nucleation fraction is high \citep{2019ApJ...878...18S}, the timescales we derive are likely upper limits.  

For the residence time in the case of evolution in a circumbinary disc we use:
\beq
t_{\rm bin,d}=1.5\times10^{-2} \,\epsilon_{0.1} \,f_{\rm Edd}^{-1} \frac{q}{(1+q)^2}\ln\left(\frac{a_i}{a_c}\right)  \, {\rm Gyr}, 
\label{eq:tbind}
\eeq
where $q\equiv M_2/M_1 \leqslant 1$ is the MBH binary mass ratio, $\epsilon_{0.1}$ is the radiative efficiency normalised to 0.1, and $f_{\rm Edd}$ is the luminosity in units of the Eddington luminosity at $t_{\rm in}+t_{\rm df}$. We follow \cite{2015MNRAS.448.3603D} in selecting $a_i=G M_{12}/2\sigma^2$ and $a_c=1.9\times10^{-3} {\rm pc} (M_{12}/10^8 \,M_{\odot})^{3/4}$ respectively as the separation when the MBHs form a binary and the separation at which emission of gravitational waves brings the MBHs to coalescence in $\sim 10^4$ yr, to obtain 
\beq 
\frac{a_i}{a_c}=\left[1.14\times10^4\left(\frac{M_{12}}{10^8 \,M_{\odot}}\right)^{1/4}\left(\frac{\sigma}{100 \, {\rm km\, s^{-1}}}\right)^2 \right]. 
\eeq

\begin{figure}
    \includegraphics[width=\columnwidth]{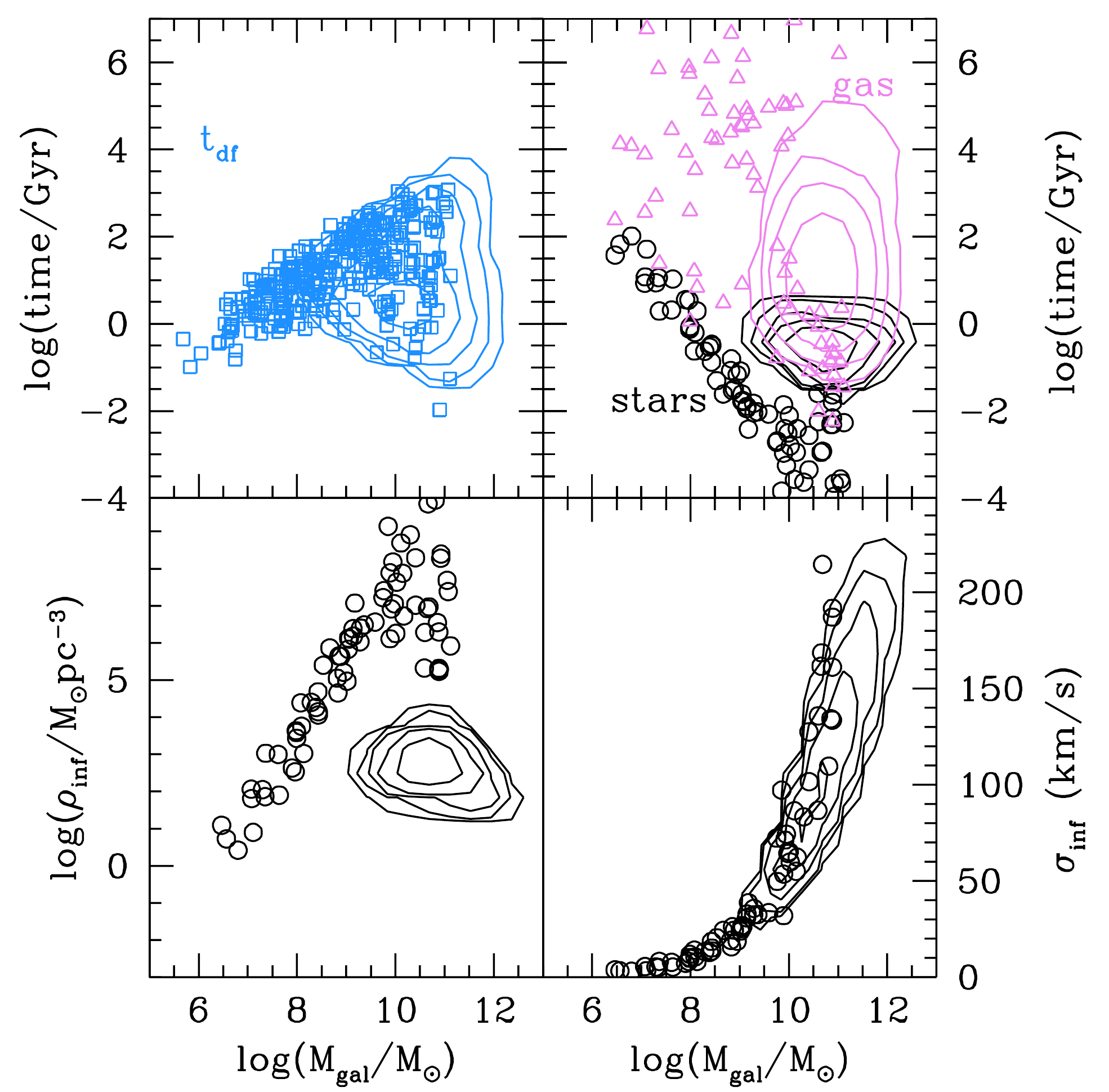} 
 	    \caption{Top: Delay timescales for \hagn~(contours) and \nh~(markers) as a function of the stellar mass of the host galaxy, for catalogue ``\rbg''.  We show the dynamical friction timescale $t_{\rm df}$ (blue), $t_{\rm bin,d}$ for gas-driven inspiral (purple) and $t_{\rm bin,h}$ for stellar hardening (black). Bottom: velocity dispersion and stellar density at the radius of the sphere of influence for \hagn~ (contours) and \nh~(circles) as a function of the stellar mass of the host galaxy, for catalogue ``\rbg''. If MBHs are surrounded by  steep stellar density cusps, then the hardening timescales are  generally shorter than one Gyr, and shorter than the timescale for migration in a gaseous disc: the bottleneck for dynamical evolution is the dynamical friction phase.   
	    }
    \label{fig:tbin_NH_HAGN}
\end{figure}

Since mass accretion onto MBHs is highly variable, we average $f_{\rm Edd}$  of the binary, i.e., of numerically merged $M_{\rm BH}$, over 50~Myr before the time when the binary forms. Recall that at this point, $t_{\rm in}+t_{\rm df}$, we have only one MBH which we consider the binary. We further assume that the evolutionary timescale is the minimum between these two: $t_{\rm bin}=\min(t_{\rm bin,h},t_{\rm bin,d})$ and we define $t_{\rm fin}=t_{\rm in}+t_{\rm df}+t_{\rm bin}$.

Our approach does not include triple MBH interactions, which can lead to fast MBH coalescences when they excite eccentricity through the Kozai-Lidov mechanism and through chaotic dynamics \citep{2018MNRAS.473.3410R,2018MNRAS.477.3910B}. Binary evolution timescales could therefore be shorter than calculated here.

In Fig. \ref{fig:tbin_NH_HAGN} we show the resulting timescales as a function of the stellar mass of the host galaxy. In general, stellar hardening in  galaxies with high central stellar densities leads to the fastest evolution, with timescales shorter than gas-driven evolution even in high-redshift galaxies. The binaries with very long $t_{\rm bin,d}$ are those with ``starving'' MBHs that have very low $f_{\rm Edd}$. From Eq.~\ref{eq:tbind}, with all other properties fixed, $t_{\rm bin,d}$ is minimum for a MBH accreting at the Eddington rate, and it gets very long when $f_{\rm Edd}$ is very low: the gas that is available for MBH accretion and for fueling a circumbinary disc comes from the same  supply in the environment of the MBH: when MBHs are starved, the gas-driven migration of a MBH binary is also stalled \citep{2015MNRAS.448.3603D}.  Binary evolution in circumbinary discs is therefore lengthened by the same processes that stunts MBH growth, for instance supernova explosions in dwarf galaxies \citep{2015MNRAS.452.1502D}. 

Stellar hardening is less effective in \hagn~for both numerical and physical reasons. Numerically, the lower spatial and mass resolution in \hagn~means that the central stellar densities are lower and therefore the timescales longer. Physically, for the most massive galaxies, the central density is generally lower than for less massive counterparts \citep{Faber1997}, and it has been argued that this density decrease is caused by the scouring during the hardening of MBH binaries \citep{Faber1997,Milosavljevic2001}  and AGN feedback \citep{2012MNRAS.422.3081M}. While the former effect is not present in the simulation, and AGN feedback does flatten the galaxy density profiles of massive galaxies in \hagn~\citep{2017MNRAS.472.2153P}.

In \nh~MBHs are lighter at a given galaxy stellar mass and furthermore galaxies are more compact, i.e., they have a smaller $R_{\rm eff}$ at a given stellar mass, therefore $\rho_{\rm inf}$ is higher and this leads to shorter hardening timescales.

\subsection{Connecting galaxy and black hole mergers}
\label{sec:galmergers} 

We construct the history of all galaxies, from their birth to the time the simulation ends ($z=0$ for HAGN and $z=0.45$ for \nh), and associate numerical MBH mergers to the galaxy mergers from which they originate.

The history of each galaxy is contained in the simulation's merger trees: we use \textsc{TreeMaker} \citep{2009A&A...506..647T} on all galaxies for \hagn~and at the final redshift for \nh. We define the main descendant of a galaxy as the one that shares the most mass in the following output. Two galaxies are defined as merged when their main descendant is the same,  and this also defines the time of the galaxy merger.

To associate MBH and galaxy mergers, for each MBH merger we select the first output where each of the MBHs in the binary can be associated to a galaxy using a threshold of $2R_{\rm eff}$ to search for a MBH in a galaxy. We then search in the merger tree to identify a galaxy merger that involves descendants of the initial galaxies, and check that  the two original MBHs were hosted in the merging galaxies. The match MBH-galaxy at the time of the galaxy merger uses a threshold of $2R_{\rm eff}$ for Horizon-AGN, but we allow the search out to $10R_{\rm eff}$ of galaxies in unpolluted haloes in \nh, with in practice 90\% of matches having distances less than $4.25R_{\rm eff}$. The reason is that, close to a merger, the galaxy centre is not a well-defined quantity because of morphological disturbances, and this is more evident in high-resolution simulations. With this procedure, for most numerical MBH mergers we identify a galaxy merger for which we know the redshift and the properties of the merging galaxies.

\section{Results}
\subsection{Black hole mergers across galaxy populations and black hole masses}

\begin{figure}
	\includegraphics[width=\columnwidth]{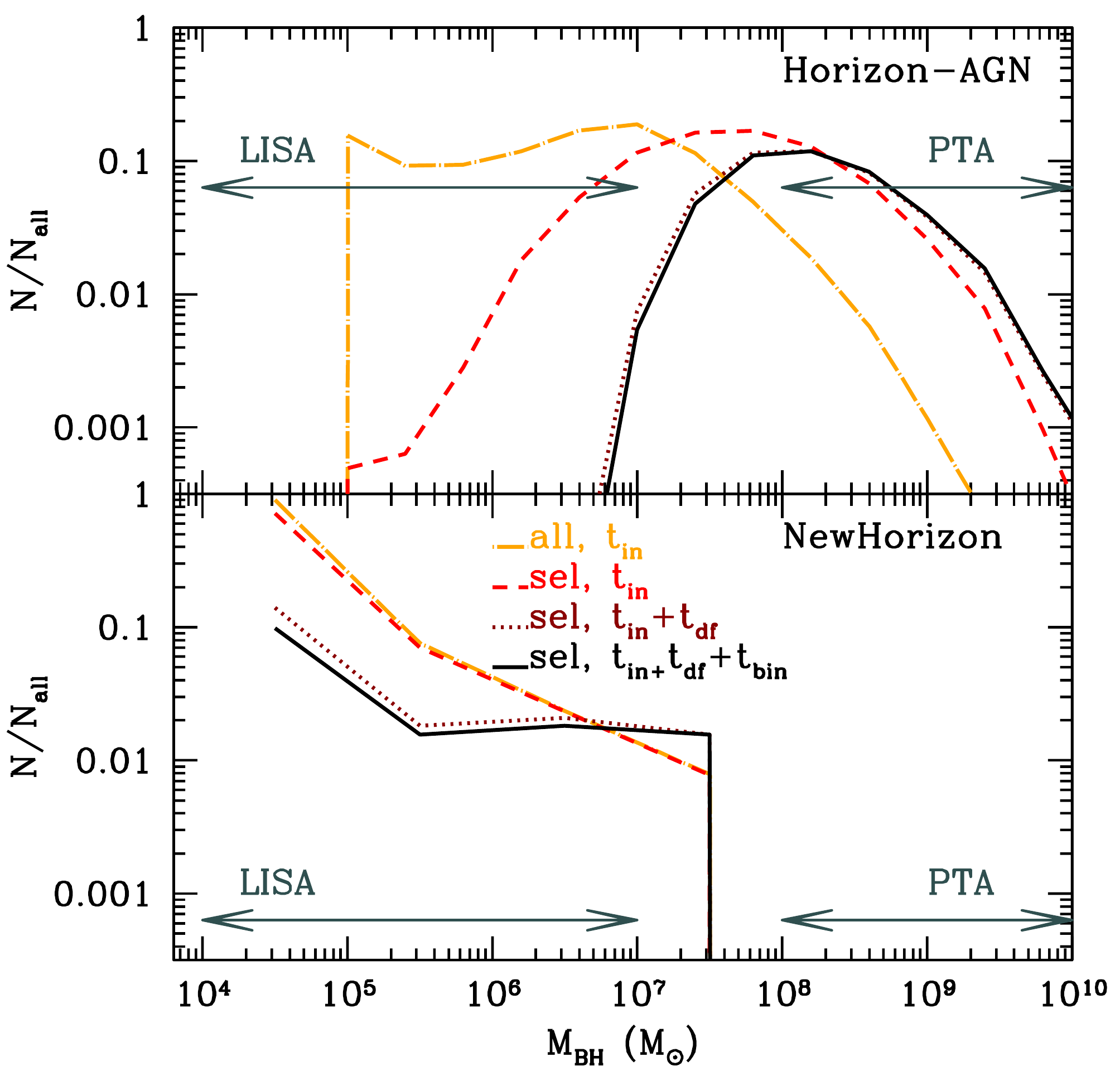}
    \vspace{-0.5cm}
    \caption{Mass of the primary MBH at the time of the numerical merger ($t_{\rm in}$), after $t_{\rm df}$ and after $t_{\rm df}+t_{\rm bin}$  for \hagn~(top) and \nh~(bottom). We show here the mergers from catalogues \all~and \rbg. Each histogram is normalised to the number of MBHs in catalogue \all~at $t_{\rm in}$. Dynamical delays shift the distributions to higher masses. We also highlight the mass ranges accessible with LISA and PTA, showing that \hagn~and \nh~are complementary. }
    \label{fig:BHmass_NH_HAGN}
\end{figure}

\hagn~and \nh~are simulations with very different characteristics:  \hagn~simulates a large volume at low spatial and mass resolution, while \nh~simulates a relatively small average volume at high spatial and mass resolution. \hagn~is therefore unable to correctly resolve the low-mass galaxies hosting the low-mass MBHs that LISA can detect: the sweet spot for LISA's sensitivity is at about $10^5-10^6 \msun$. Conversely, \nh~does not contain any massive galaxies hosting MBHs massive enough to contribute to the PTA signal, which is dominated by MBHs with mass $>10^8 \msun$ \citep{PTA2008}.  This is exemplified in Fig.~\ref{fig:BHmass_NH_HAGN} where we show the distribution of the MBH masses in the two simulations. We note that $M_2$ is not followed self-consistently after the numerical merger: at any time after the numerical merger, $M_{\rm BH}$ is the sum of $M_1$ and $M_2$ plus any mass accreted after that time.

In \hagn~the distribution of masses in the full (\all) sample, shown as an orange dash-dotted curve,  is relatively flat from the seed mass out to $10^7 \msun$, and drops afterwards. When we remove spurious merger events, a large fraction of low-mass MBH mergers disappear (\rbg, red dashed cuve). When we look at $t_{\rm in}+t_{\rm df}$ (dark red dotted curve), MBHs with mass $<10^6 \msun$ virtually disappear, while  at the high mass end there are two competing effects: on the one hand some of largest MBHs are removed from the sample, because the time between $t_{\rm in}$ and $z=0$ is too short, on the other hand MBHs grow during $t_{\rm df}$, extending the distribution to larger masses. Finally, we look at the mass distribution at $t_{\rm in}+t_{\rm df}+t_{\rm bin}$, where we find a similar behaviour. 
In \nh~we find a more limited loss of MBHs at the low-mass end because fewer spurious MBH mergers occur in the outskirts of galaxies, but the overall trends are similar. 

\begin{figure}
	\includegraphics[width=\columnwidth]{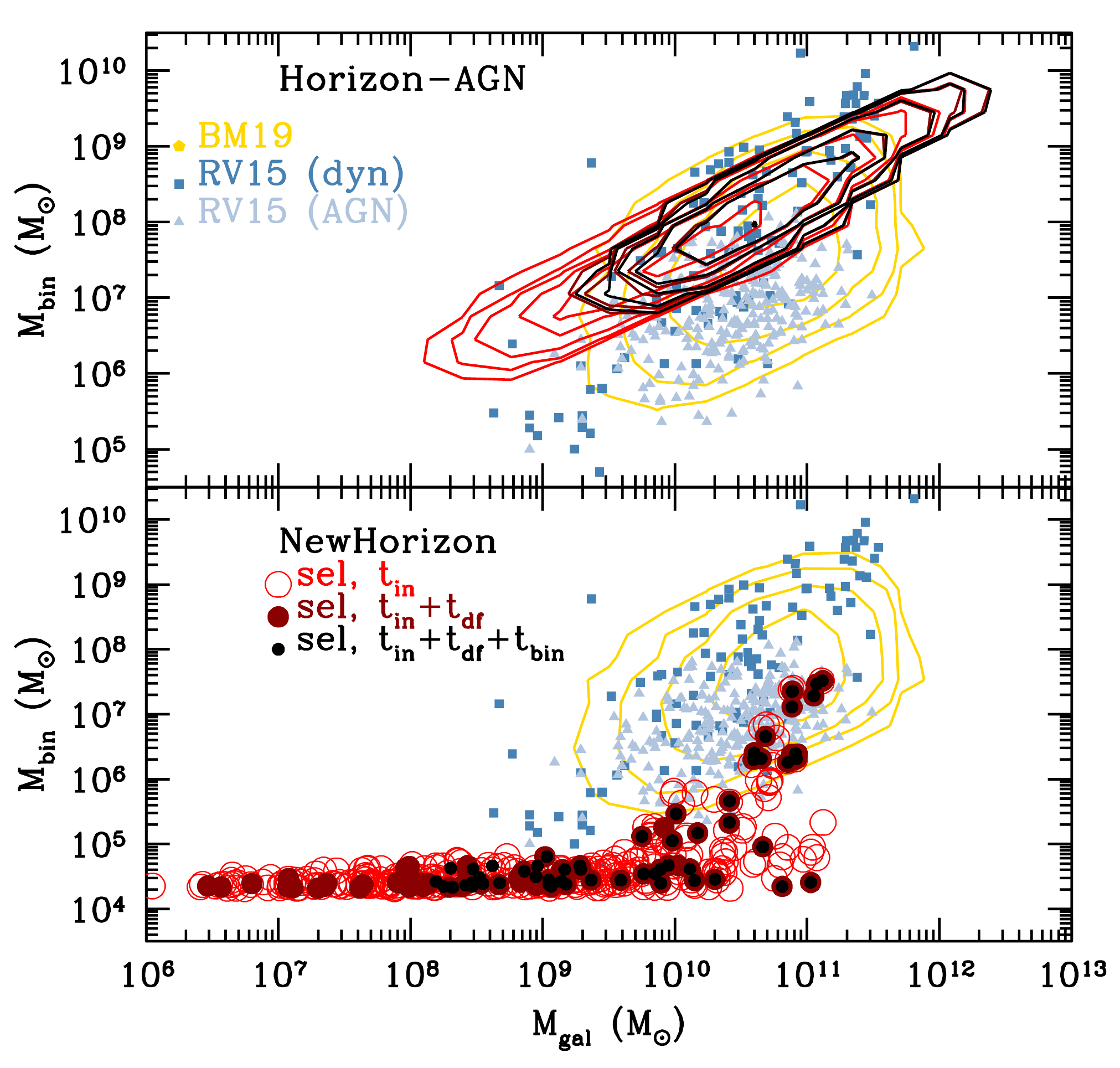}
    \vspace{-0.5cm}
    \caption{Mass of MBH binaries versus the stellar mass of the host galaxies at the time of the numerical merger, after $t_{\rm df}$ and after $t_{\rm df}+t_{\rm tbin}$ for \hagn~(top, four logarithmically spaced contours) and \nh~(bottom, individual galaxies and MBHs are shown) for catalogue \rbg. The observational results are for single MBHs at $z\sim 0-0.3$ \citep{2015ApJ...813...82R,2019MNRAS.487.3404B}.
     }
    \label{fig:galBHmass_NH_HAGN}
\end{figure}

Fig.~\ref{fig:galBHmass_NH_HAGN} shows the relation between the masses of MBH binaries and the total stellar mass of their host galaxies at different times: the numerical merger, $t_{\rm in}$, after the dynamical friction phase, $t_{\rm in}+t_{\rm df}$, and after the binary evolution,  $t_{\rm in}+t_{\rm df}+t_{\rm bin}$. The observational samples are from \cite{2015ApJ...813...82R} and \cite{2019MNRAS.487.3404B}. \cite{2015ApJ...813...82R} include both quiescent galaxies with dynamically measured MBH masses, typically hosted in spheroids, and active galaxies with single-epoch MBH masses estimates using broad AGN lines (type I AGN),  hosted in both spheroids and disc galaxies. \cite{2019MNRAS.487.3404B} propose a novel way to estimate MBH masses from narrow AGN lines (type II AGN) and apply it to AGN at $z<0.3$.

In \hagn~at given galaxy mass the binary masses are consistent with the masses of single MBHs in the simulation, and show the same trends as single MBHs, i.e., they reproduce well the data at the high-mass end but they overpredict MBH masses at the low mass end, and the distribution has a smaller scatter than in observations (see \cite{2016MNRAS.460.2979V} for a complete discussion). In \nh, MBH growth is delayed by SN feedback \citep{2015MNRAS.452.1502D} and MBH growth picks up only in galaxies more massive than $5\times10^{9} \msun$ \citep{2019arXiv190408431D} when major gas inflows allow MBHs to grow significantly above their seed masses. MBHs in the most massive galaxies by the end of the simulation, $z=0.45$ have masses in agreement with observational samples, but there are no MBHs with mass $>10^6$ in galaxies with stellar mass $10^9-10^{10} \msun$. We note that this delayed MBH growth leads to a better match of the theoretical AGN luminosity function with observations in \nh~(Beckmann et al. in prep.) than in \hagn~\citep{2016MNRAS.460.2979V}, where it is overestimated.

\begin{figure}
	\includegraphics[width=\columnwidth]{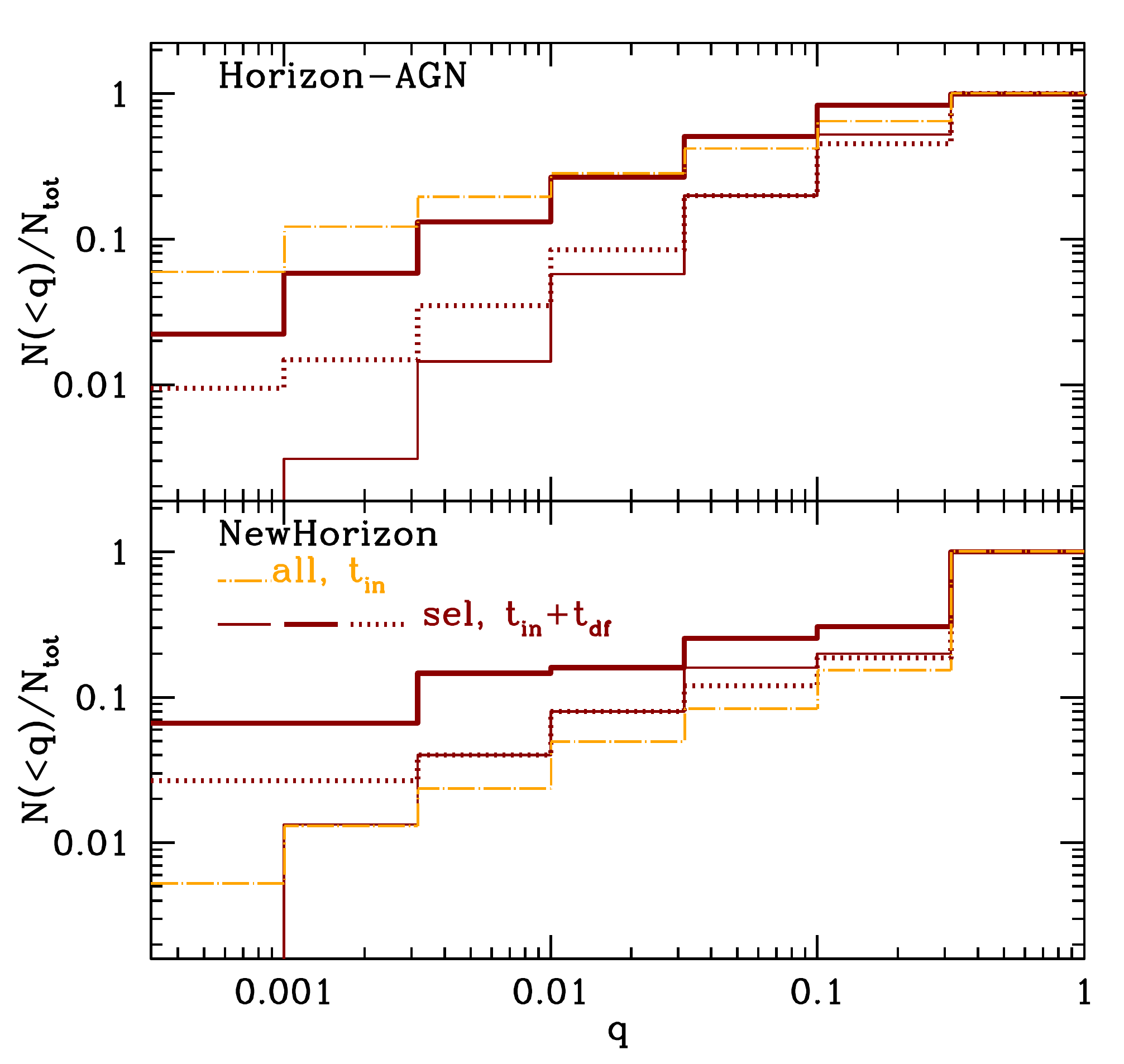}
    \vspace{-0.5cm}
    \caption{Cumulative distribution of mass ratios of merging MBHs. The orange dot-dashed histogram is $q=M_2/M_1$ for \all~numerical mergers, the thin solid histogram for MBHs at the time of binary formation ($t_{\rm in}+t_{\rm df}$) keeping the same ratio as at $t_{\rm in}$ and the thick solid histogram for the same binaries, but calculating $q$ as $q=M_2/(M_{\rm BH}(t_{\rm in}+t_{\rm df})-M_2)$, while the thick dotted histogram assumes that there has been a coup and all the mass growth occurred on $M_2$, i.e. $q=M_1/(M_{\rm BH}(t_{\rm in}+t_{\rm df})-M_1)$. Each histogram is normalised to unity. Despite the difficulty in defining a mass ratio after the MBHs are numerically merged, generally the distributions are skewed towards high mass ratio values, $q=0.1-1$.}
    \label{fig:mass_ratio_HAGN_NH}
\end{figure}

\begin{figure*}
    \begin{overpic}[width=0.5\textwidth]{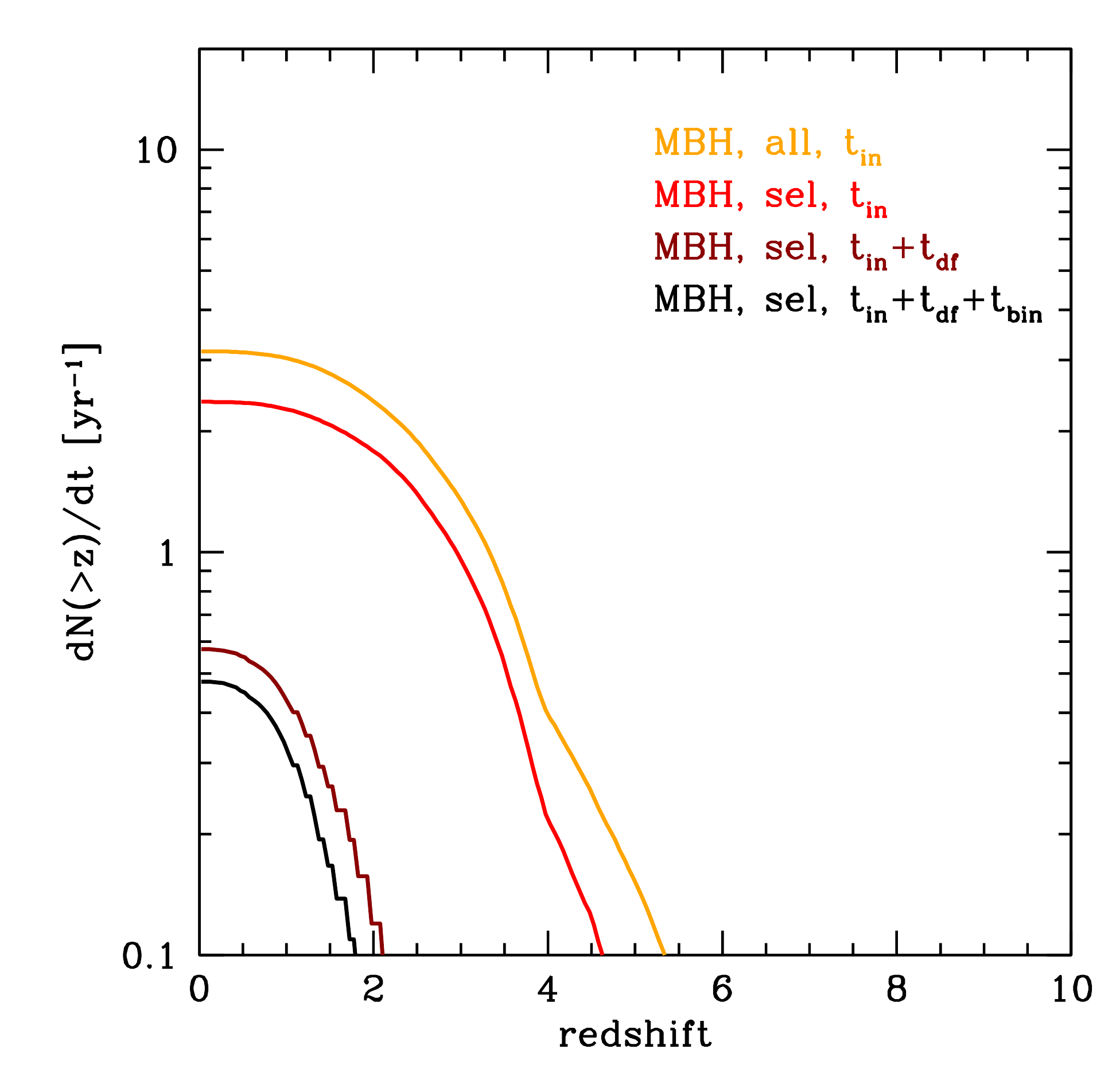} 
	\put(24,89){\hagn}
    \end{overpic}\hfill
    \begin{overpic}[width=0.5\textwidth]{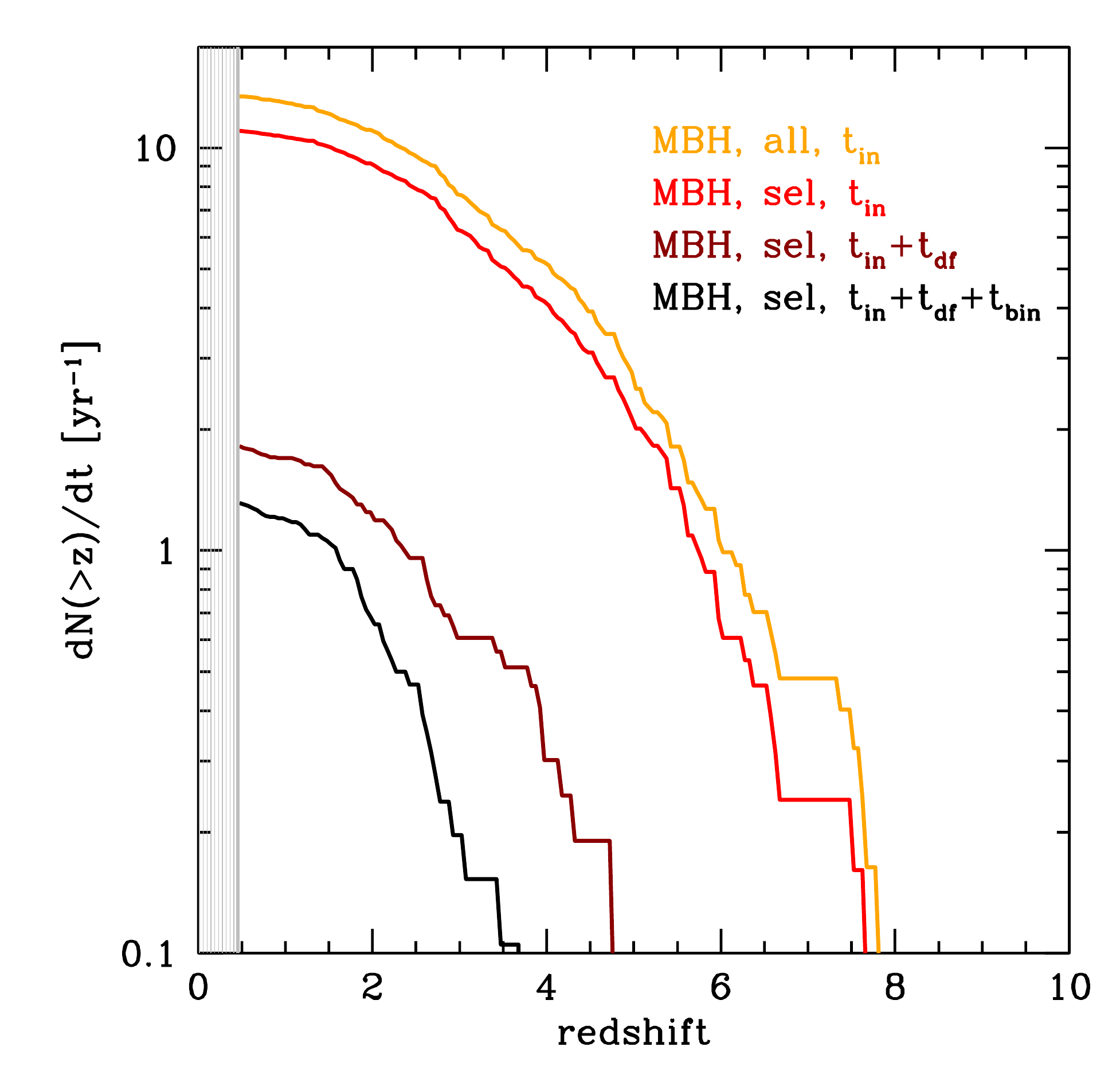}
	\put(24,89){\nh}
    \end{overpic}
    \vspace{-0.5cm}
\caption{MBH observable cumulative merger rate in \hagn~(left panel) and \nh~(right panel, the grey shaded area marks the region not covered by the simulation, which ends at $z=0.45$. Note the difference in y-axes for the two simulations. Although \nh~has a much smaller volume, the resulting merger rate is higher because the merger history of dwarf galaxies is resolved. As noted in Fig.~\ref{fig:BHmass_NH_HAGN}, the masses of merging MBHs in the two simulations are very different: up to $10^7 \msun$ for \nh~and generally above to $10^7 \msun$ for \hagn, therefore the two merger rates are relevant for LISA and PTA respectively. Since \nh\ simulates an average volume of the Universe, it does not contain any high-bias region where mergers are enhanced, thus it gives a lower limit to the merger rate.}
    \label{fig:BH_mergrate}
\end{figure*}

The typical mass ratio of merging binaries is an important information in preparation for LISA: at the time of writing, waveforms are available for mass ratios $q\sim 1$ and $q\ll1$, but the range $q=10^{-1}-10^{-6}$ falls in between those that can be studied with numerical relativity,  post-Newtonian techniques and gravitational self-force \citep{2019CQGra..36n3001B}. Understanding if these mass ratios are statistically significant is therefore of great interest to the waveform community. Unfortunately, the mass ratio of the merging binaries is well defined only at $t_{\rm in}$: at that point in the simulation $M_1$ and $M_2$ are merged and therefore subsequently the two masses cannot be followed separately although we artificially consider the two MBHs as still under dynamical evolution. We can however follow the total mass evolution of the binary, by tracking the numerically merged black hole.

For reference we define the mass ratios at $t>t_{\rm in}$ in three different ways: either we consider that the mass ratio has not changed, i.e., that $M_2$ has grown at the same pace as $M_1$, or we consider that $M_2$ has not changed and all mass growth occurred on $M_1$, or that there has been a coup \citep{VW2014} with a swap between  $M_1$ and $M_2$ so that all subsequent growth occurred on $M_2$. The truth could be anywhere in between and also outside this range, but we refrain from trying to speculate further, although we note that during both the dynamical friction phase and the evolution in a circumbinary disc $M_2$ is expected to grow faster than $M_1$ \citep{Callegari2011,2012ApJ...755...51N,2015MNRAS.447.2123C,2015MNRAS.447L..80F}.

The mass ratios are shown in Fig. \ref{fig:mass_ratio_HAGN_NH} for the full population at $t_{\rm in}$ (orange dot-dashed) and for the MBHs that form a binary at $t_{\rm in}+t_{\rm df}$. Adding dynamical delays shifts the \hagn~distribution to larger mass ratios, while it has the opposite effect in \nh. In \hagn~this is due to the long dynamical friction time when $M_2$ is small, while in \nh, where a large fraction of MBHs can bind, the effect is due to the growth of $M_\mathrm{BH}$ with respect to $M_2$ (thick lines) and to mergers with initial $q\approx 1$  involving two MBHs with
$M_\mathrm{BH} \approx 10^4 \msun$, which have long merging timescales. In summary, a large fraction of the binaries should have mass ratio between 0.1 and 1, but a tail at $q<0.1$ cannot be excluded.

\subsection{Massive black hole merger rates}

The MBH merger rate from the two simulations, defined as the rate measurable from an observer on Earth over the whole sky \citep{Haehnelt1994} is shown in Fig. \ref{fig:BH_mergrate}. First, we show how removing ``spurious'' mergers affects the results. We compare the numerical MBH merger rates for catalogues \all~and \rbg: in \hagn~almost 80\% of mergers occur outside $2R_{\rm eff}$. This is not the case for \nh, although in both simulations most mergers within $2R_{\rm eff}$ actually occur between half and twice the galaxy effective radius. 

We stress that \nh~provides only a lower limit to the merger rate, since it simulates an average region of the Universe and biased regions provide a significant contribution to the merger rate even taking into account for their rarity \citep{Sesanaetal2005}. Although the simulations should not be combined since they are not self-similar in sub-grid physics, the merger rate of MBHs for LISA should be higher than the sum of the rates of MBHs with mass $<10^7 \msun$ from \nh\ and \hagn: \nh\ does not have biased regions, and \hagn\ does not resolve low-mass galaxies.   Furthermore, as noted in section~\ref{sec:selection}, the lack of a direct  implementation of dynamical friction from stars and dark matter in the simulations -- only dynamical friction from gas is included on-the-fly -- also goes in the direction of reducing the number of MBH mergers, since with additional friction the MBHs would have a smoother dynamics, bind more easily, and numerical mergers would be facilitated.

Including dynamical delays in post-processing can severely reduce the raw merger rate from numerical mergers \citep{2020MNRAS.491.2301K}. Adding delays  shifts the merger rate to lower redshift, with a peak at $z\sim1-2$. This is favorable for electromagnetic counterpart searches, since the counterparts will be brighter and with better sky localization from LISA \citep{McGee2020}. We postpone a complete study of the counterparts to a future study. 

The  merger rate is generally lower than that predicted by semi-analytical models, mostly because only few semi-analytical models include dynamical modelling of MBH orbital decay and binary evolution \citep{VHM,2012MNRAS.423.2533B,2019MNRAS.486.4044B}, and even in these cases the early evolution is normally approximated by integrated dynamical friction timescales such as Eq.~\ref{eq:tdf} that do not capture the more erratic and stochastic behaviour seen in simulations  \citep{2019MNRAS.486..101P,2020arXiv200502409B}. Our study includes integrated dynamical friction timescales in post-processing, but before the numerical merger dynamics is calculated directly in the simulation:   MBHs respond to inhomogeneous, time-varying conditions and the direct implementation of dynamical friction, although only from gas, allows us to account for evolution of the MBH mass and of the environment on-the-fly. Recently \cite{2020arXiv200603065B} include  in a semi-analytical additional kpc-scale delays before formation of binaries, and show that the merger rate is then greatly reduced.

Furthermore, the highest merger rate in semi-analytical models is predicted when including the mergers of the remnants of Population~III stars in very high redshift galaxies \citep{GW3,2016PhRvD..93b4003K,2018MNRAS.481.3278R,2019MNRAS.486.2336D}, which we do not treat in our simulations \citep[see][for a careful comparison between simulations and semi-analytical models]{2020MNRAS.491.2301K}.

The predicted merger rate from \hagn~is also somewhat lower than in EAGLE or Illustris \citep{2016MNRAS.463..870S,2020MNRAS.491.2301K} because in our simulations MBHs are not constantly repositioned at the center of the potential minimum, which makes MBHs merge immediately after galaxy mergers: even adding delays in postprocessing simulations with MBH repositioning facilitate  mergers.    In both  \hagn~and \nh~the large scale dynamics is driven by the explicit use of a drag force, and we include delays in post-processing only below resolution.

The most important point in the comparison between \hagn~and \nh, however, is that a small, high-resolution simulation like \nh~predicts higher merger rates than a large, low-resolution simulation like \hagn. The reason is the ability to track mergers of low-mass galaxies hosting MBHs. Although the occupation fraction of MBHs in galaxies is predicted to decrease with galaxy mass \citep{VLN2008,2010MNRAS.408.1139V}, observationally it seems to be between 0.1 and 1 in galaxies with mass $\sim10^9 \msun$ at $z=0$ \citep{2015ApJ...799...98M,2017ApJ...842..131S} and between 0.5 and 1 when considering local galaxies within 4 Mpc and mass $\sim10^9-10^{10} \msun$ and published MBH dynamical masses or limits \citep{2018ApJ...858..118N}. We refer to \cite{2019arXiv191109678G} for a review. At $z=0.45$ the occupation fraction of MBHs within $2R_{\rm eff}$ in \nh~is 0.1 at $M_{\rm gal}=10^6 \msun$ and it reaches unity at $M_{\rm gal}=10^9 \msun$, while it becomes about 2, i.e., there are on average two MBHs within a galaxy, at  $M_{\rm gal}=10^{10.5} \msun$. Tracing MBHs in dwarf galaxies is crucial.

In the context of LISA's science, simulations like \hagn, EAGLE or Illustris, which do not resolve dwarf galaxies, will underestimate the merger rate of LISA's MBHs. A simulation like Romulus \citep{2017MNRAS.470.1121T}, with volume and resolution intermediate between \hagn~and \nh, and also seeding MBHs in low-mass galaxies like \nh, is a good compromise. Ideally, we would like a simulation with the resolution of \nh~and the volume of \hagn~run to $z=0$: unfortunately this is currently computationally unfeasible.

\begin{figure}
    \includegraphics[width=\columnwidth]{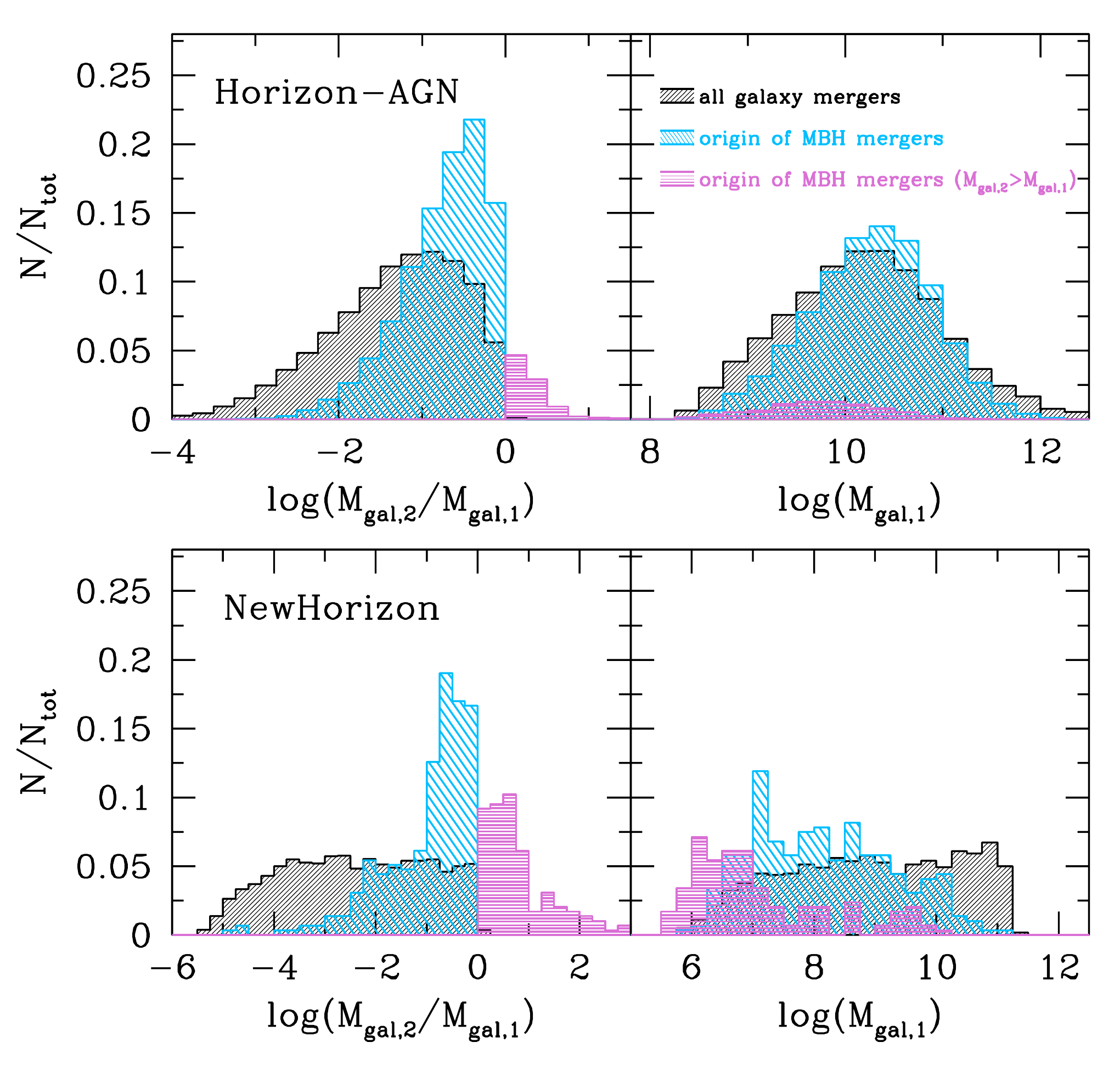}
    \vspace{-0.5cm}
   \caption{Properties of galaxy mergers in  \hagn~(top two panels) and \nh~(bottom two panels). In black we show the distribution for all galaxy mergers in the simulation, and in blue the distribution for the galaxy mergers that lead to MBH mergers in catalogue \rbg. $M_{\rm gal,1}$ is the stellar mass of the most massive of the two merging galaxies, so the distribution is truncated at a mass ratio equal to unity. Each histogram is normalised to the total number of objects in the respective distribution. We show in violet the distribution for the cases where the least massive MBH is hosted in the most massive galaxy of the pair, normalised to the total number of galaxy mergers leading to MBH mergers. The masses of the galaxies belonging to this subset are shown in violet in the right panels.  The distribution of galaxy mergers sourcing MBH mergers is skewed towards larger mass ratios with respect to the full underlying galaxy merger population. }
    \label{fig:galmergprops}
\end{figure}

\begin{figure}
	\includegraphics[width=\columnwidth]{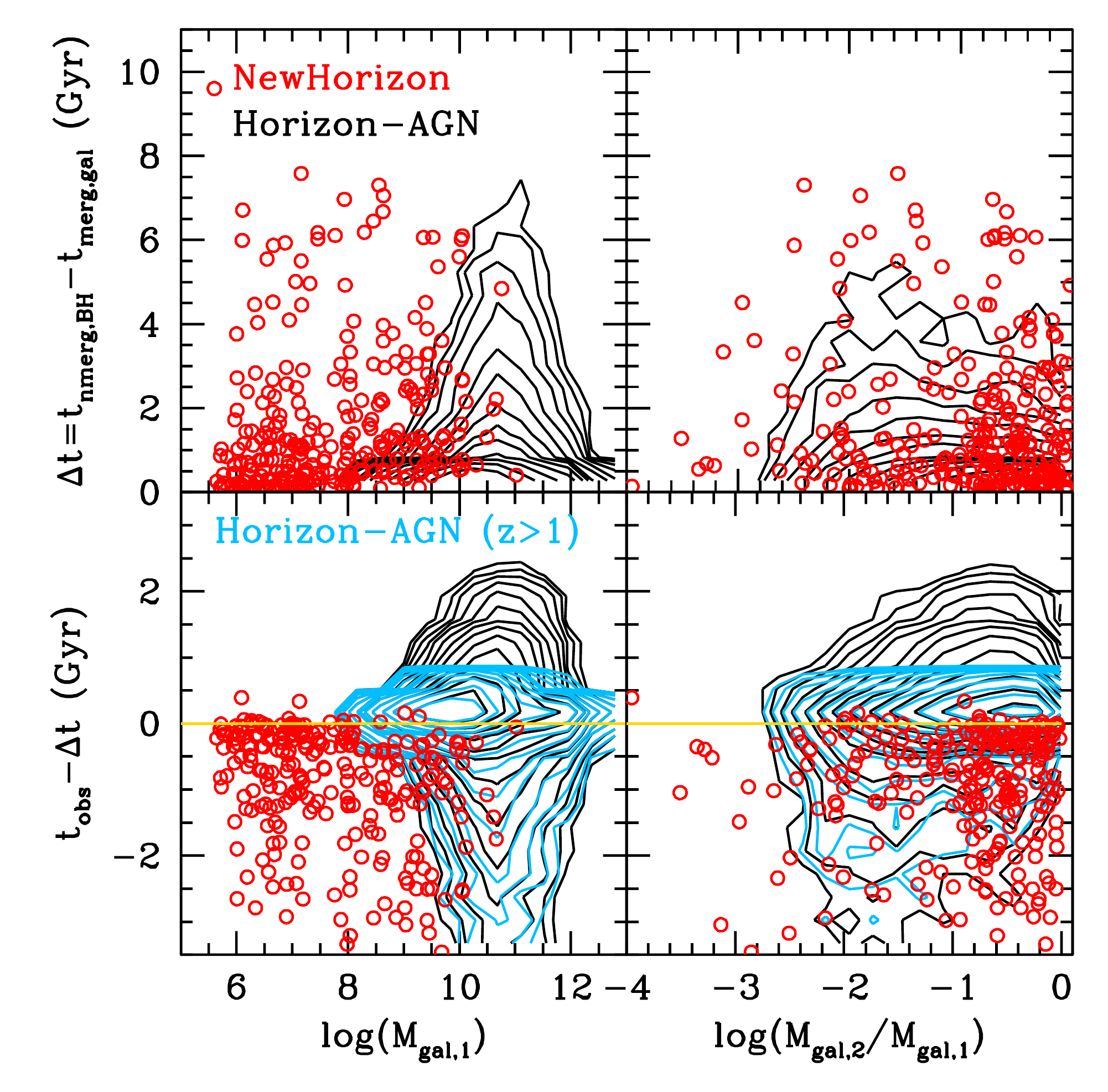}
    \caption{ Top: delay between galaxy merger and MBH numerical merger in \hagn~(black contours) and \nh~(red circles) as a function of mass of the primary galaxy, $M_{gal,1}$ (left), and galaxy mass ratio (right) for catalogue \rbg.  \nh~has been analysed down to $z=0.45$, when the age of the Universe is almost 9~Gyr, which is why the delay times are always shorter than $\sim 8$~Gyr. Bottom: difference between  the time during which one would observe the galaxy pair as actively merging and the time elapsed from the galaxy merger and the MBH numerical merger, a lower limit to the effective time of the MBH merger. For \hagn~we  show 20 logarithmic spaced contours  in black,  with the light blue contours for mergers at $z>1$. All MBHs below the yellow horizontal line will merge after the galaxies are in an observable merger phase. The MBH mergers most relevant for LISA, in galaxies with mass $<10^{10} \msun$, occur after any sign of the galaxy merger that generated the MBH merger has disappeared in terms of the structure or kinematics of the galaxy. }
    \label{fig:galBHdelay}
\end{figure}

\subsection{Which galaxy mergers lead to black hole mergers?}

\begin{figure*}
	\begin{overpic}[width=\columnwidth]{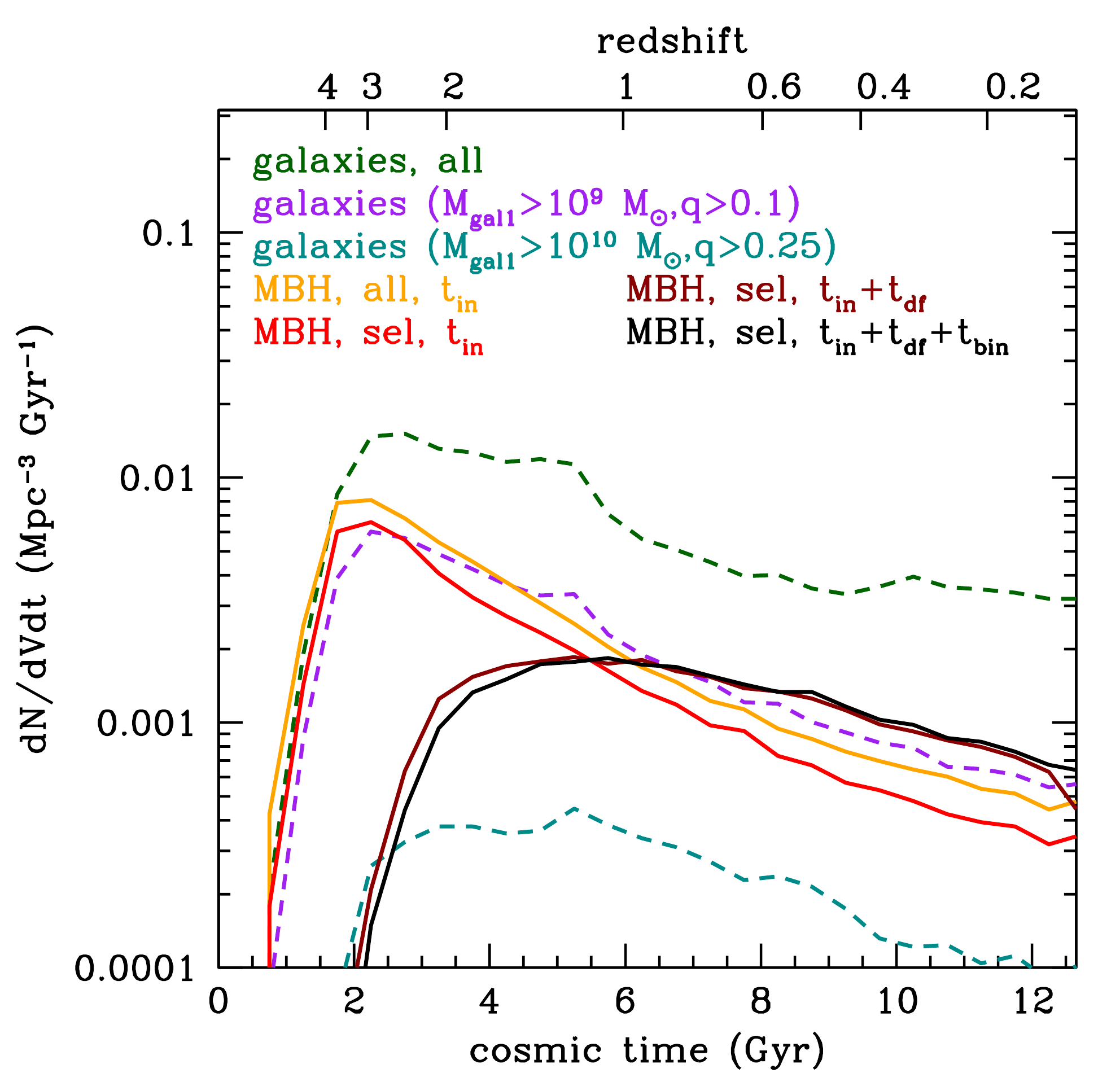}
	\put(72,83){\hagn}
    \end{overpic}
	\begin{overpic}[width=\columnwidth]{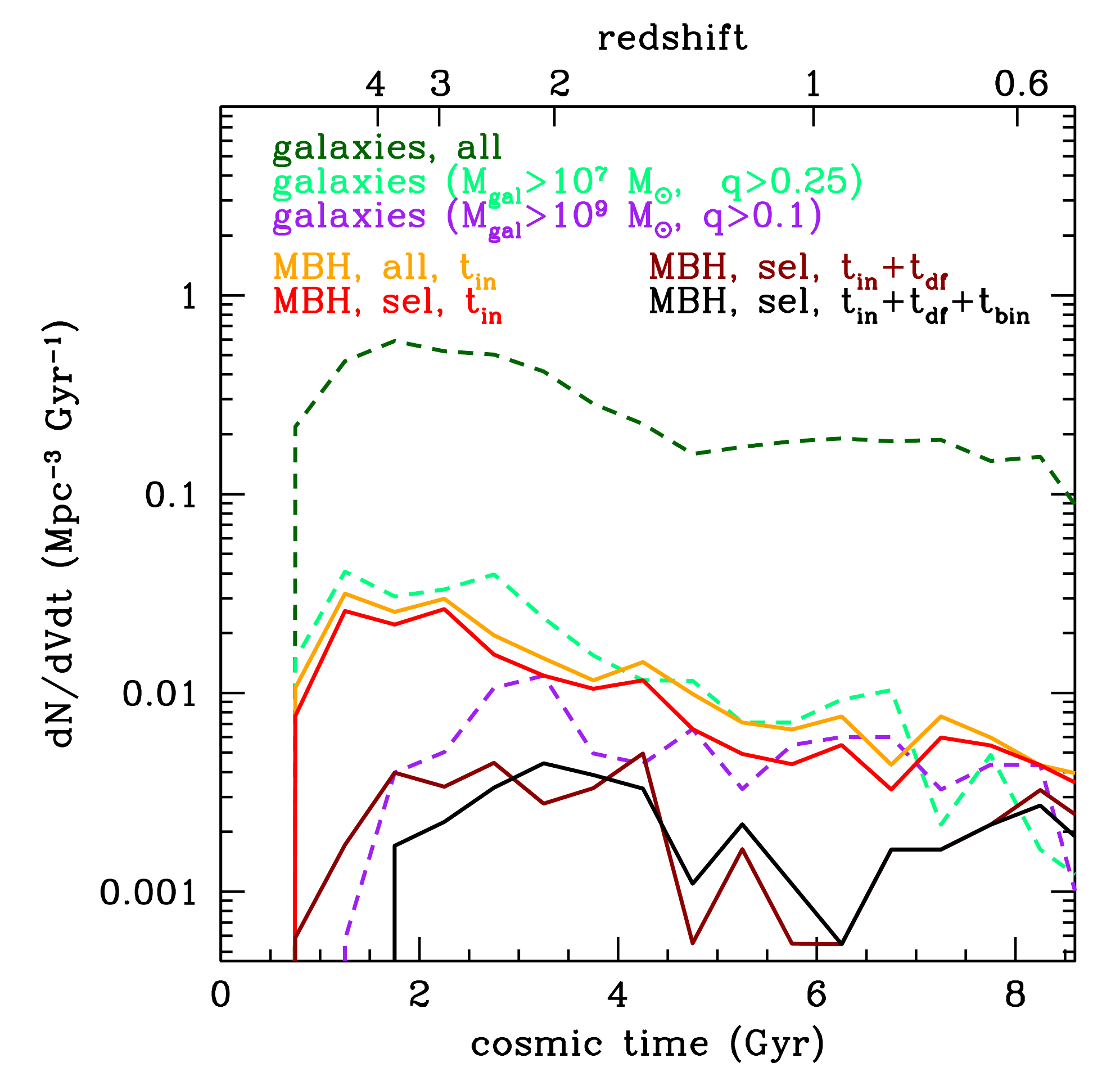}
	\put(72,83){\nh}
    \end{overpic}
    \caption{Comparison between the galaxy and MBH merger rate for  \hagn~(left) and \nh~(right, note the different y-axes). This is shown as intrinsic merger rate, not as the observable merger rate. For the galaxy merger rate we show some examples of stellar mass and mass ratio cuts. The MBH ``numerical'' merger rate differs from the galaxy merger rate in normalization, while including dynamical delays changes also the shape of the distribution.
    }
    \label{fig:comp_gal_BH_mergrate}
\end{figure*}

Beyond semi-analytical models and simulations, another technique often used to estimate the MBH merger rate is through the galaxy merger rate, assuming scaling relations between galaxies and MBHs \citep{2013MNRAS.433L...1S,2016ApJ...826...11S,2019MNRAS.488..401C}. We test this approach here, by relating MBH mergers to the galaxy merger they originated from. In our general picture, at the beginning of a galaxy merger, each MBH sits at the centre of the merging galaxy, and for mass ratios of the merging galaxies $>1:4$ we expect that the two MBHs end up in the centre of the merger in a bound binary \citep{BBR1980,Callegari2009,Svanwas2012,2017MNRAS.471.3646P}. In the case of low-mass MBHs, however, this picture may not capture the full behaviour. Low-mass MBHs may not reside in the centre of their host galaxy \citep{2019MNRAS.482.2913B} and/or their dynamical evolution can be subject to disturbances \citep{2019MNRAS.486..101P} preventing the formation of the binary even in mergers between similar mass galaxies.  

We compare the whole distribution of galaxy stellar masses and mass ratios to that of galaxy mergers that lead to MBH mergers in Fig. \ref{fig:galmergprops}.  Additional parameters play a role, e.g., the compactness of the satellite galaxy \citep{2018MNRAS.475.4967T}, but these quantities are hard to measure in observations, therefore we focus here only on masses and mass ratios, since these are ``macroscopic'' quantities than can be measured in observations, although even measuring a galaxy stellar mass mass requires some modelling from imaging and spectroscopy.

We consider here ``numerical MBH mergers'', while the results when we include dynamical friction are discussed further down. We define $M_{\rm gal,1}$ as the most massive of the two merging galaxies, but in $\sim$10\% of cases the least massive MBH is hosted in the most massive galaxy of the pair (violet histograms in Fig. \ref{fig:galmergprops}). Each histogram is normalised to the total number of objects in the respective catalogue, and in \hagn~the galaxy mergers identified as sources of a MBH merger represent 16\% of the total number of mergers for catalogue \rbg, and 37\% of catalogue \all; the fractions for \nh~are even smaller (3.5\% and 4\% respectively): not all galaxy mergers lead to a MBH merger. The reason is twofold: the occupation fraction of MBHs is below unity in low-mass galaxies \citep{Volonteri2008}, so the probability of a merger between two galaxies each hosting a MBH is also much lower than unity (as a first approximation this probability scales as the occupation fraction squared, although biased regions experience an enhanced number of mergers). Furthermore, even if galaxies merge, the MBHs can be stalled at large separations in the galaxy even for major mergers \citep{2018MNRAS.475.4967T}. 

As expected, galaxy mergers that generate MBH mergers are a biased sample with respect to the general merging population, specifically they have a higher mass ratio compared to the full distribution: in \hagn~the mean mass ratio is 0.05 for the general population, and 0.17 for \rbg, 
for \nh~they are respectively 0.004 and 0.1 so that even minor mergers, i.e., with mass ratio $<1:4$ contribute to sourcing MBH mergers, and in fact they represent more than 50\% of the sample. 

In \hagn, the masses of galaxies sourcing MBH mergers are somewhat larger than those of the full merging population. The reason is that the occupation fraction of MBHs is larger in more massive galaxies \citep[for \hagn~see Figures~9 and~10 in][for \nh~the trend is similar at  $M_{\rm gal}>10^{8.5} \msun$ and follow the same slope at lower galaxy mass, reaching $\sim 0.1-0.3$ at $M_{\rm gal}=10^6 \msun$. Note that both in \hagn~and in \nh~some galaxies host multiple MBHs]{2016MNRAS.460.2979V}. In the case of \nh, the increase in galaxy mass seems to disappear, but this is because most mergers involving galaxies with mass $>10^{10} \msun$ are very minor mergers. If we restrict the mass distribution to mergers with mass ratio $>0.1$ the masses of primary galaxies in mergers leading to MBH mergers are 0.5 dex larger than the mean galaxy masses of the global merging population (the same is true if we perform the same mass ratio cut in \hagn). 

When we require that the dynamical friction timescales added in post-processing bind the MBHs within the Hubble time at $z=0$, the mean mass ratios increase to 0.22 and 0.14 for \hagn~and \nh, while the galaxy masses decrease slightly. Dynamical friction timescales are shorter the larger the mass of the infalling MBH, and therefore the more massive its host galaxy, $M_{\rm gal,2}$, in general, while they are shorter the smaller $M_{\rm gal,1}$ is, via $\sigma$ in Eq. \ref{eq:tdf}. Furthermore, as mergers involving massive galaxies happen at later times, there is less time to complete the dynamical friction phase by $z=0$.

In Fig.~\ref{fig:galBHdelay} we show the time delay between when galaxies merge and when their MBHs are numerically merged, $\Delta t=t_\mathrm{nmerg,BH}-t_\mathrm{merg,gal}$, for the \rbg~selection. We recall that the time of the merger is defined as when two galaxies have the same main descendant, i.e., only one galaxy is identified by the halo finder. These delays do not include any of the timescales added in post-processing, therefore they are lower limits to the time when MBHs actually merge. Overall, there is a wide range of delays at any given galaxy mass or mass ratio and there is no simple fit/trend to describe the delays.

Typically when mergers occur, the signatures of these events in the remnant morphologies, structures, and kinematics will last for around 1 - 1.5 Gyr, at most \citep{2006ApJ...638..686C}.  This is however for the central parts of the remnant galaxy which is most readily visible, and often the only parts visible.  Outer tidal features may persist for some time, and shells for far longer  \citep{2017Galax...5...34P} but both of these are very difficult to observe as they only exist in low surface brightness light, which even for nearby galaxies is not easily seen even when using extremely deep exposures and performing careful sky subtraction.  The structural peculiarities in the central parts of these remnant mergers will last for a maximum of 2 Gyr or so. If we use the merger time-scale from \cite{2017MNRAS.468..207S} and \cite{2019ApJ...876..110D} of 2.4 Gyr $(1+z)^{-2}$ we obtain the result shown in the bottom panel of Fig.~\ref{fig:galBHdelay} demonstrating that this merger time has elapsed before most galaxies will merge their black holes.  This 2.4~Gyr time, scaled by redshift, is the pair timescale to go from 30~kpc to effectively the start of the merger process.  This is therefore the time for two galaxies to merge, but not necessarily the time it would take to coalesce into a single system.  After this 2.4~Gyr time, scaled by redshift, there will be some time afterwards when the galaxy is still morphologically and structurally distorted, before it dynamically relaxes.  However, this timescale would be about $1.5  (1+z)^{-2} \,{\rm Gyr}$ \citep[e.g.,][]{2017MNRAS.468..207S}.  This time-scale is shorter at higher redshifts, meaning that the merger signatures will disappear even faster at earlier times. 

 Overall, taking $t_{\rm obs}=2.4  (1+z)^{-2} \,{\rm Gyr}$, the values shown in Fig.~\ref{fig:galBHdelay} are upper limits to the time we could identify these systems as mergers through pair selection or a morphological approach such as the CAS  (concentration C, asymmetry A, and clumpiness S) method for ongoing mergers, or the use of visual morphologies or machine learning techniques to identify peculiar structures \citep[e.g.,][]{2003ApJS..147....1C}. For the simulation, we consider  the time elapsed from the galaxy merger and the MBH numerical merger, $\Delta t$, which sets a lower limit on MBH merger timescales as we do not include any of the delays discussed in Section~\ref{sec:delays}.  The choices have been very conservative: for MBHs we do not include all the time delays after the numerical merger, and therefore underestimate the time between galaxy and MBH merger. For galaxies, the $2.4(1+z)^{-2}$ Gyr timescale is the time to start galaxy coalescence from when they are separated by 30 kpc. The $1.5 (1+z)^{-2}$ Gyr timescale is the post-merger time over which discernible features are observable. The time we consider for the MBHs is the post-merger phase, therefore it should be compared to the $1.5 (1+z)^{-2}$ Gyr timescale, but we considered the $2.4 (1+z)^{-2}$ Gyr timescale to be more conservative and have a more robust result. 

Using this criterion, all MBHs below the yellow horizontal line will merge after the host galaxy can be observed to be in interaction. The difference between the two simulations can be ascribed to two main reasons: (i) \hagn~includes mergers at $z<1$ for which  $t_{\rm obs}$ is longer, and (ii) at a fixed galaxy mass MBHs are lighter in \nh, therefore the orbital decay longer: the simulations include on-the-fly dynamical friction following \citep{Ostriker1999}, where the deceleration depends linearly on the mass of the MBH.

The result of this is that if the LISA sources are identified in a galaxy it is very likely that the system will be dynamically and morphologically relaxed given the long time-period between the galaxy merger and the merger of the central black holes, as long as no further galaxy merger intervenes during $\Delta t$. Massive black holes in massive galaxies at low redshift, i.e. those relevant for PTA, are more likely to be found in galaxies that show signs of interaction caused by the same galaxy merger that supplied the MBH for the MBH merger. MBHs hosted in relatively small galaxies at high redshift, which are most relevant for LISA, will merge in galaxies where the signs of the interaction that generated the merger have been erased. However, at $z>1$ the galaxy merger rate is very high, and a galaxy can experience one or more galaxy mergers during $\Delta t$, as can be seen in the examples in  Section~\ref{sec:galev}. It is important, however, to realise that the disturbed morphology of the galaxy is not caused by the same galaxy merger from which the MBH merger originated: without taking this effect into account  risks mis-associating galaxy and MBH mergers, which would lead to  incorrectly inferring short merger timescales for MBHs.

\begin{figure}
    \includegraphics[width=\columnwidth]{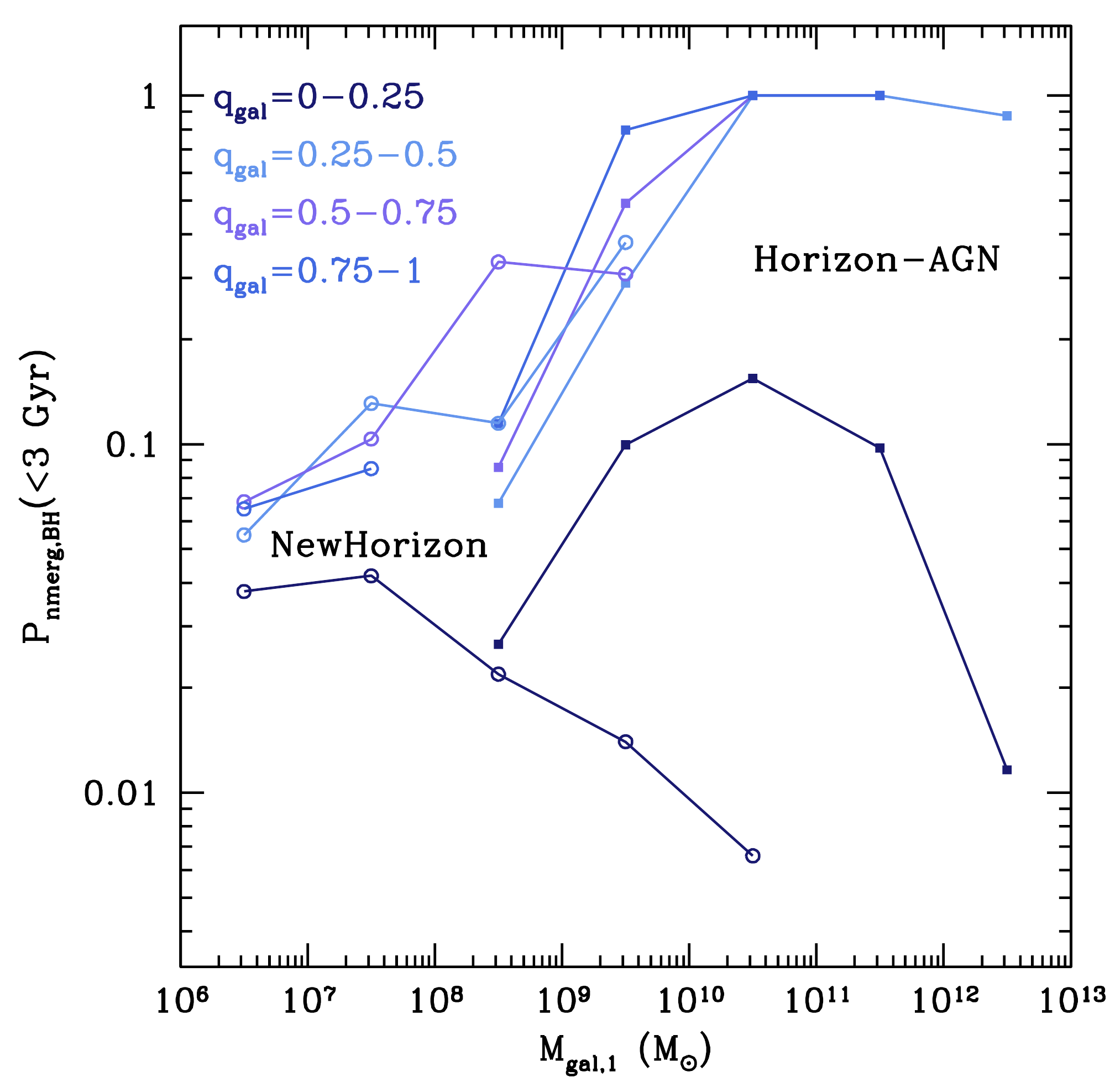}
    \vspace{-0.5cm}
    \caption{Probability that a galaxy merger with a primary galaxy stellar mass $M_{\rm gal,1}$ and mass ratio $q_{\rm gal}=M_{\rm gal,2}/M_{\rm gal,1}\leqslant 1$ is followed by a MBH merger within 3~Gyr. This probability takes into account that not all galaxy mergers lead to a MBH merger, see Fig.~\ref{fig:galmergprops} and related discussion in the text.
    }
    \label{fig:prob_merger_time}
\end{figure}

\begin{figure*}
\centering
    \includegraphics[width=0.32\textwidth]{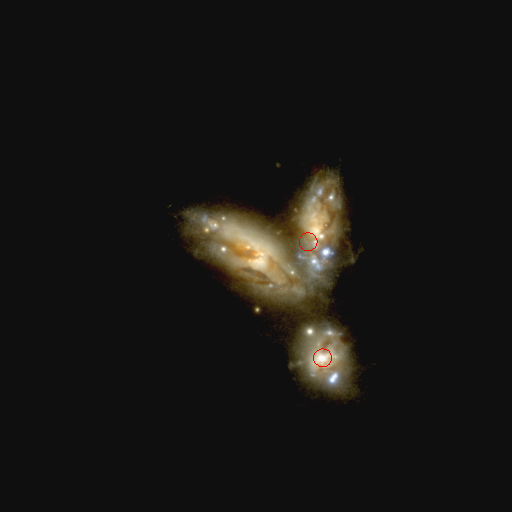}
    \includegraphics[width=0.32\textwidth]{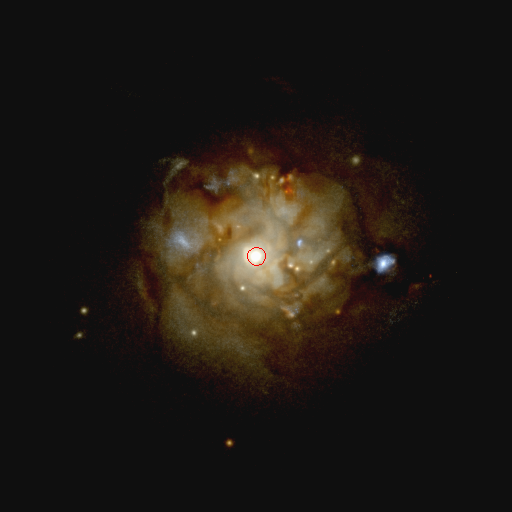}
    \includegraphics[width=0.32\textwidth]{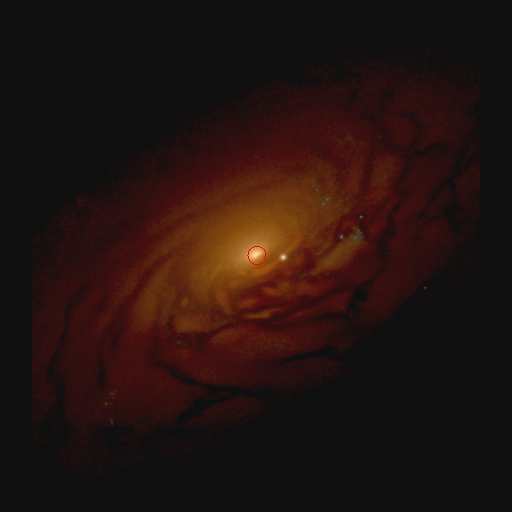}
\caption{False colours ugr image of the evolutionary sequence for the MBH merger 936+41 in \nh. The red circles mark the position of the MBH(s). From left to right: the galaxy merger ($z=3.18$, $t=2.00$~Gyr), the time of the MBH numerical merger ($z=2.61$, $t=2.56$~Gyr), and at the end of the simulation, $z=0.45$, the dynamical friction timescale would end at $z=0.01$. Image sizes: 50~kpc, 50~kpc, 20~kpc.}
\label{fig:936_41}
\end{figure*}

\begin{figure*}
\centering
    \includegraphics[width=0.32\textwidth]{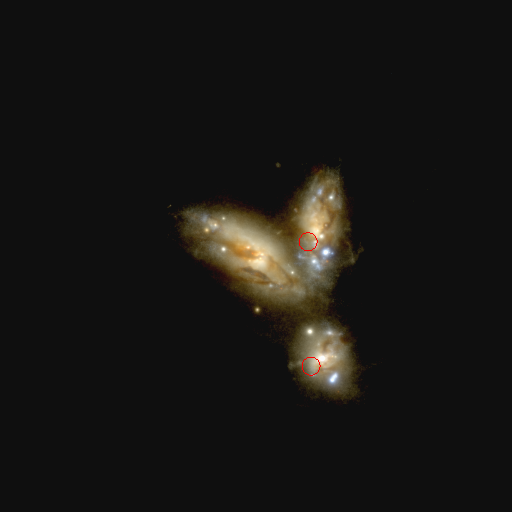}
    \includegraphics[width=0.32\textwidth]{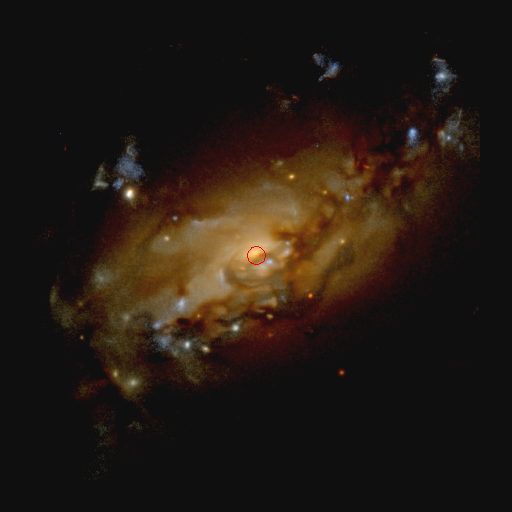}
    \includegraphics[width=0.32\textwidth]{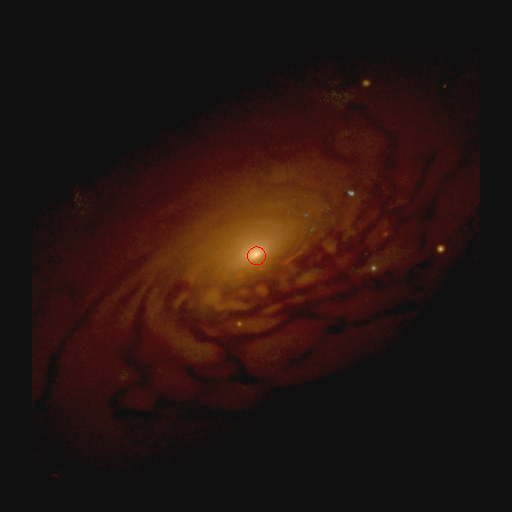}
\caption{False colours ugr image of the evolutionary sequence for the MBH merger 936+96 in \nh. The red circles mark the position of the MBH(s). From left to right: the galaxy merger ($z=3.18$, $t=2.00$~Gyr; this is the same merger shown in Fig.~\ref{fig:936_41}, left), the time of the MBH numerical merger ($z=1.38$, $t=4.67$~Gyr), and after the dynamical friction timescale has elapsed $z=0.60$, $t=8.02$~Gyr). Image sizes: 50 kpc, 20 kpc, 20 kpc.}
\label{fig:936_96}
\end{figure*}

We conclude this section by comparing the galaxy\footnote{For a general discussion on galaxy mergers in simulations, see, e.g.,  \cite{2015MNRAS.449...49R}. We only note there that the galaxy merger rate of \nh~and \hagn~is in good agreement with observations \citep{2019ApJ...876..110D}.} and MBH merger rate in Fig. \ref{fig:comp_gal_BH_mergrate}. As already noted, the total number of galaxy mergers is larger than the total number of MBH mergers, because not all galaxies host MBHs, and not all galaxy mergers lead to a MBH-MBH merger. If we consider only numerical mergers, i.e., without including delays in post-processing, the change between the galaxy and MBH merger rate is mostly in the normalization, because at least some of the delays shown in Fig. \ref{fig:galBHdelay} are short. Adding additional time to the delay between the numerical merger and the formation of the binary or the merger of the binary causes a shift in both normalization and shape, i.e., MBH mergers are shifted to later cosmic times. We show some reference cuts in mass and mass ratio to galaxy mergers inspired by the distributions shown in Fig. \ref{fig:galmergprops}. Given that delay times are not easily connected to basic observable galaxy properties (mass, mass ratio), converting a galaxy merger rate into a MBH merger rate is non-trivial. Although the full physical picture of linking MBH and galaxy mergers is complex, it is generally possible to infer statistical MBH merger rates from galaxy merger rates, but the cuts in mass and mass ratio should be tuned to the particular MBH properties of interest.

As Fig.~\ref{fig:prob_merger_time} shows, the probability for two galaxies to merge and then to contain a MBH  numerical merger within 3 Gyr is highest for the higher stellar mass systems, and appears to drop in the \hagn~simulation at lower masses, although this drop is less steep for the \nh~simulation, where low-mass galaxies are better resolved and have a higher MBH occupation fraction.   This probability takes into account that not all galaxy mergers lead to a MBH merger, explaining why although most MBHs have $\Delta t=t_\mathrm{nmerg,BH}-t_\mathrm{merg,gal}<3$~Gyr in Fig.~\ref{fig:galBHdelay}, the probability is generally $<1$, except at the high-mass end for mass ratios $>0.25$. What this plot shows is that we are more likely to find a MBH merger in more massive galaxies.  Observationally, this is important if we want to trace the merger history of MBHs from examining the merger history of galaxies (e.g., Conselice et al. 2020 in prep).  Interestingly, this plot also shows that there is a higher probability for MBH mergers to occur for major mergers where the $q_{\rm gal}=M_{gal,2}/M_{gal,1} \leqslant 1$ ratio is larger than for more minor mergers.  For shorter time periods, this decreases at lower galaxy masses and lower mass ratio of galaxy mergers, and for longer time periods there is a slight increase, with differences being more pronounced for \nh.

\begin{figure*}
\centering
    \includegraphics[width=0.32\textwidth]{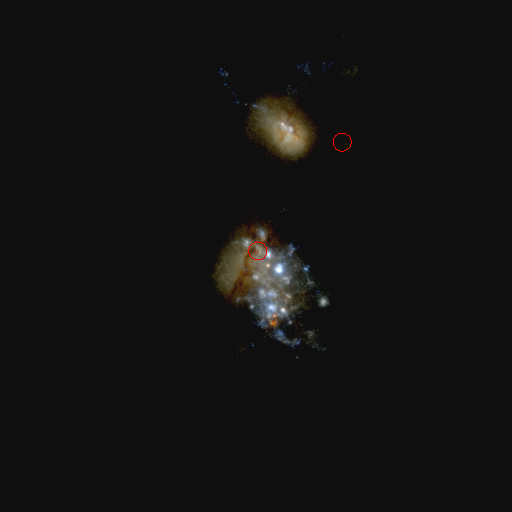}
    \includegraphics[width=0.32\textwidth]{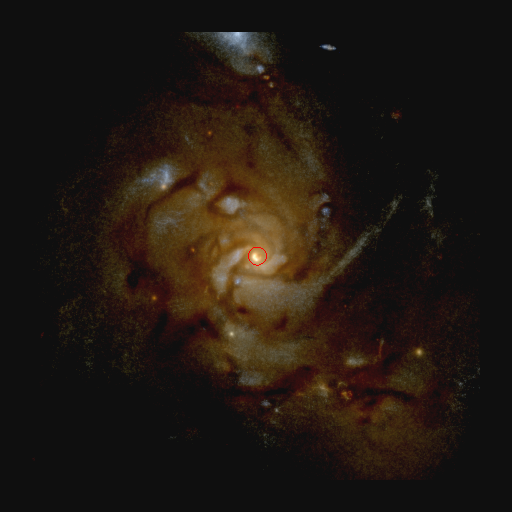}
    \includegraphics[width=0.32\textwidth]{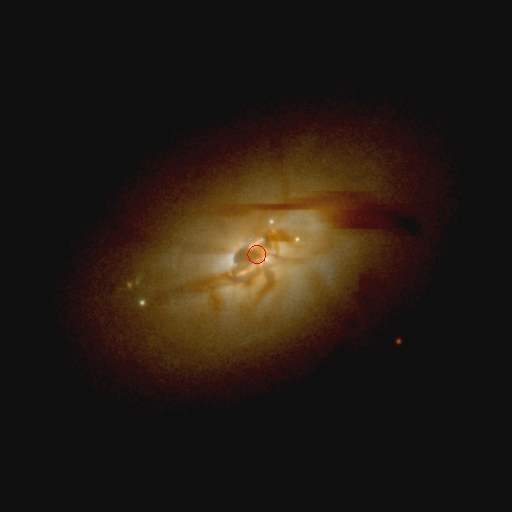}
\caption{False colours ugr image of the evolutionary sequence for the MBH merger 796+9 in \nh. The red circles mark the position of the MBH(s). From left to right: the galaxy merger ($z=3.76$, $t=1.64$~Gyr), the time of the MBH numerical merger ($z=1.58$, $t=4.17$~Gyr), after the dynamical friction timescale has elapsed ($z=1.19$, $t=5.29$~Gyr). Image sizes: all 20 kpc.}
\label{fig:796_9}
\end{figure*}

\begin{figure*}
\centering
    \includegraphics[width=0.32\textwidth]{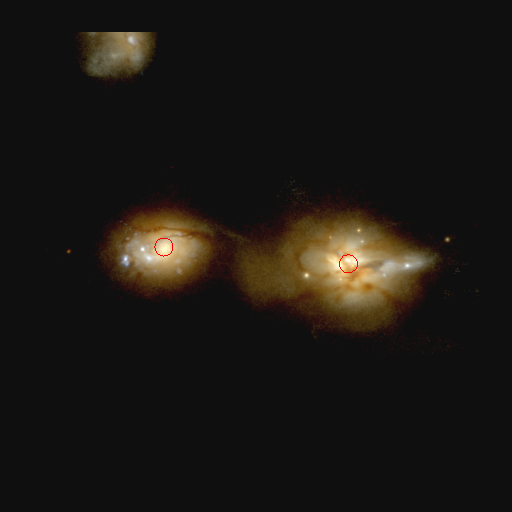}
    \includegraphics[width=0.32\textwidth]{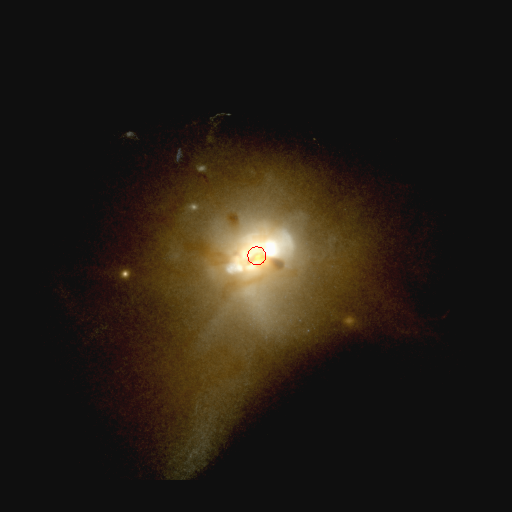}
    \includegraphics[width=0.32\textwidth]{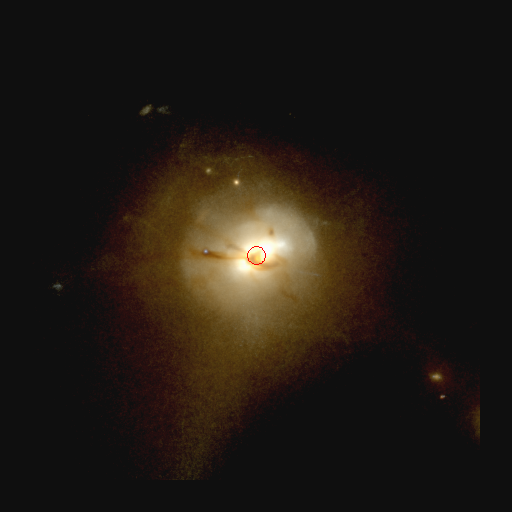}
\caption{False colours ugr image of the evolutionary sequence for the MBH merger 166+656 in \nh. The red circles mark the position of the MBH(s). From left to right: the galaxy merger ($z=2.24$, $t=2.95$~Gyr), the time of the MBH numerical merger ($z=1.93$, $t=3.48$~Gyr), after the dynamical friction timescale has elapsed ($z=1.90$, $t=3.53$~Gyr). Image sizes: all 50 kpc.}
\label{fig:166_656}
\end{figure*}

\begin{figure*}
\centering
    \includegraphics[width=0.32\textwidth]{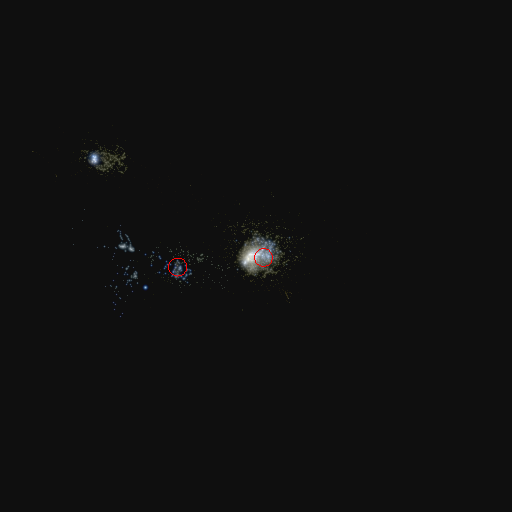}
    \includegraphics[width=0.32\textwidth]{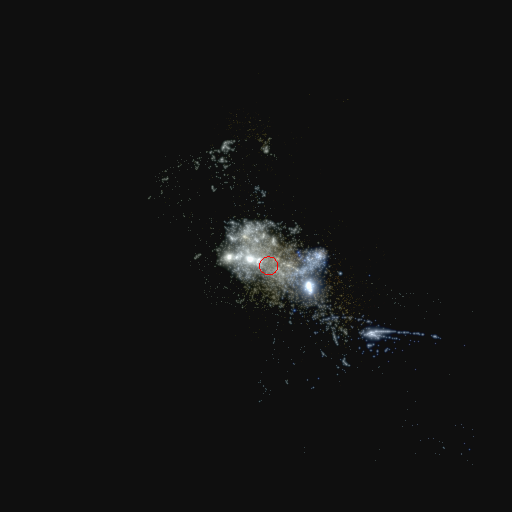}
    \includegraphics[width=0.32\textwidth]{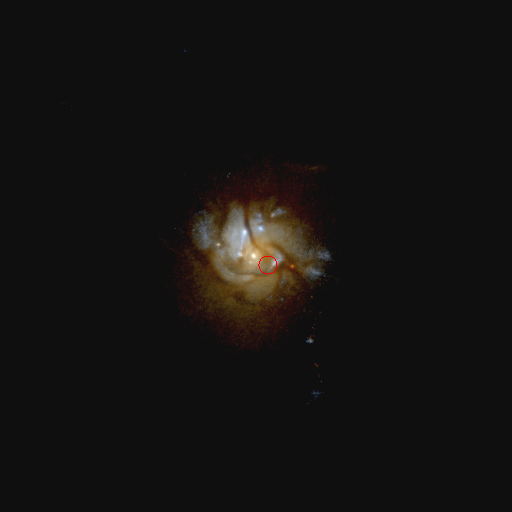}
\caption{False colours ugr image of the evolutionary sequence for the MBH merger 455+156 in \nh. The red circles mark the position of the MBH(s). From left to right: the galaxy merger ($z=7.37$, $t=0.68$~Gyr), the time of the MBH numerical merger ($z=6.39$, $t=0.88$~Gyr), after the dynamical friction timescale has elapsed ($z=2.43$, $t=2.76$~Gyr). Image sizes: 10 kpc, 10 kpc, 20 kpc.}
\label{fig:455_156}
\end{figure*}

\subsection{Galaxy evolution during black hole dynamical evolution}
\label{sec:galev}

While the observability of electromagnetic counterparts, from the MBHs themselves or from the host galaxies, will be the subject of a future work, let us study some representative cases in \nh, to highlight the importance of dynamical delays in determining what type of galaxies host MBH mergers. 
In the following figures, galaxies are represented with false-colour maps of their surface brightness through rest-frame u-g-r filters including the absorption by dust.

In the first example of a MBH merger, of MBHs `936' and `41' (Fig. \ref{fig:936_41}), which originated from a merger between two galaxies with mass $2\times10^9 \msun$ and $10^{10} \msun$ at $z=3.18$, the MBHs are numerically merged at $z=2.6$, with an ensuing dynamical friction timescale of about 10.8~Gyr, meaning that it finishes at $z\sim 0.01$, beyond the end time of the simulation. Disturbed features in the host galaxy, with mass $6\times 10^{10} \msun$, at $z=1.19$ are not related to the memory of the initial merger at $z=3.18$ but to a merger with mass ratio 0.73 at $z=1.22$. 

The second example is another merger involving MBH `936', this time with MBH `96' (Fig. \ref{fig:936_96}). It started with the same galaxy merger as `936' and `41', with MBHs `41' and `96' in the same galaxy at that time, but the MBH numerical merger happens much later, at $z=1.38$, when $2.67$~Gyr have elapsed from the initial galaxy merger. The galaxy, however, has experienced another 0.2 mass ratio merger at $z=1.39$. Dynamical friction ends at $z=0.6$, when the galaxy is almost $7\times 10^{10} \msun$.

 In the third MBH merger, of MBH `796' and MBH `9' (Fig. \ref{fig:796_9}), the initial galaxy merger took place at $z=3.76$, with a mass ratio of 0.77, between two galaxies with mass $\sim 10^9\msun$. The two MBHs are numerically merged at $z=1.58$, $\sim 2.53$~Gyr later, when the galaxy is experiencing a new merger, with a galaxy 5 times lighter. The ensuing dynamical friction and binary evolution timescales end at $z=1.19$, corresponding to a cosmic time of $\sim 5.29$~Gyr. During the latter phase of dynamical evolution, the galaxy has grown to $4\times10^{10} \msun$ and experienced four additional mergers with mass ratio $>0.1$. 

The fourth case, MBHs `166' and `656' (Fig.~\ref{fig:166_656}), starts with the merger of two galaxies with mass ratio 0.32 at $z=2.24$, where the least massive galaxy has a mass of $10^{10} \msun$. The numerical merger happens 0.5~Gyr later, when the galaxy has not relaxed yet from the merger, and our estimates for subsquent delays are very short, less than 0.1~Myr each, since the MBHs numerically merge very close to the centre of a galaxy with high stellar and gas density. This is a genuine case where the host galaxy is disturbed because of the galaxy merger from which the MBH merger originated. 

The fifth example, of MBH `455' merging with MBH `156' (Fig. \ref{fig:455_156}), originated in a very minor galaxy merger at $z=7.4$ with the most massive galaxy weighing $2.5\times10^7 \msun$ and the least massive $10^6 \msun$. The two MBHs are numerically merged at $z=6.4$, $\sim 0.18$~Gyr later, and despite the short intervening time the host galaxy has experienced a further merger with a mass ratio of 0.18.  The dynamical friction timescale is $\sim 1.87$~Gyr, ending at $z=2.43$. When the MBHs form a binary, the host galaxy has increased its mass to $7\times10^{9} \msun$. By this time the host galaxy has experienced two additional mergers with mass ratio $>0.1$ and is on its way to another major galaxy merger, but none of these are the origin of the MBH merger in question.

We end therefore with a word of caution on looking for the host galaxies of LISA MBH mergers among galaxies with signs of interactions: a galaxy could be involved in a merger at the time of a MBH binary coalescence, but it would generally be a random coincidence, due to the elevated merger rate of galaxies at high redshift.

\section{Conclusions}
\label{sec:conclusion}

In this paper we have compared the predictions for MBH mergers in two simulations, one large-volume low-resolution, the other small-volume, high-resolution. \hagn~is one of the largest hydrodynamical simulations to date run with full galaxy formation physics and produces a MBH population in good agreement with observations at mass $>10^7 \msun$. \nh~is a zoom within \hagn, run with very high mass and spatial resolution, able to resolve dwarf galaxies, and contains $\sim$ 17 galaxies with masses larger than $10^{10}\msun$ at $z=0.45$, but no MBHs with masses $>2\times 10^7 \msun$ at the same redshift. The two simulations can be considered as prototypical for studying PTA's and LISA's MBHs resepctively.

We select MBHs merged in the simulation and add delays in post-processing to describe dynamical processes occurring on sub-resolution scales. We then connect MBH mergers to the galaxy mergers from which they originated to study the connection between galaxy and MBH mergers. We summarise our main results here. 

 \begin{itemize}
 
 \item In low-resolution simulations, modelling delays requires significant extrapolations since MBHs are merged when they are at very large separation and galaxy properties are smoothed over larger scales.

 \item  If the galaxy central stellar density is high, stellar hardening is the leading process shrinking binaries, more effective than  migration in circumbinary gas discs.
 
 \item Including sub-grid dynamical delays causes a loss of events and shifts the peak of the MBH merger rate to $z\sim1-2$. 
 
 \item Considering macroscopic properties that can be obtained from observations (mass, mass ratio), galaxy mergers leading to MBH mergers are a biased sample of the general merging galaxy population and the time delay between galaxy and MBH merger has a large scatter at fixed merger properties. This complicates converting a galaxy merger rate into a MBH merger rate.  

\item  The time delay between galaxy and MBH merger is generally longer than the time over which the galaxies would be classified as ``in interaction'' or ``disturbed''. The hosts of LISA's MBHs may however appear to be disturbed because of further intervening mergers, caused by the high merger rate of high-z galaxies. 
 
 \item The merger rate estimated from a small high-resolution simulation is larger than the one from a much larger low-resolution simulation because the latter misses the low-mass galaxies that dominate the galaxy merger rate. This is especially important for the low-mass MBHs relevant for LISA.
 
\end{itemize}

Simulations are starting to resolve the masses of galaxies hosting MBHs of interest for LISA in sufficient large numbers for (small) statistical studies, but this improvement in volume/resolution is not sufficient to obtain realistic merger rates. It must be accompanied by the inclusion of appropriate sub-grid physics to track as faithfully as possible MBH dynamics down to the smallest scales. This includes not pinning MBHs to galaxy centres, as that removes all the dynamical evolution and causes artificially early MBH mergers \citep{2015MNRAS.451.1868T}. Unresolved dynamical friction from dark matter, stars and gas must be included, and, especially when investigating LISA's MBHs, care must be also put into keeping a good ratio of MBH seed mass to dark matter/stellar/gas particle to avoid spurious oscillations, especially for MBH seeding procedures motivated by MBH formation models, which in some cases can have seed masses as low as 100 $\msun$.  

We note also that \nh~is seeded with MBHs with mass $10^4 \msun$, which have been shown by \cite{2019MNRAS.486..101P} to have erratic dynamics at high redshift, leading them to stall in mass growth making them hardly able to bind in binaries. Higher mass seeds suffer less from this effect, therefore if in the ``real'' Universe MBH seeds were typically more massive than $10^4 \msun$, but formed with the same number density, the merger rate would be higher. 

This leads to the inference of a statistical argument on MBH seeds: if LISA does not detect any high-redshift mergers, it means that there are not many seeds with mass $\sim 10^5 \msun$ hosted in galaxies with conspicuous stellar and gas content allowing binary hardening and migration. This result is the opposite of many semi-analytical models that do not include the erratic seed behaviour \citep{GW3,2016PhRvD..93b4003K,2018MNRAS.481.3278R,2019MNRAS.486.2336D}, since light seeds are more common than heavy seeds, i.e., they have a larger number density \citep{2016MNRAS.457.3356V,2019MNRAS.486.2336D}. In a nutshell, light MBH seeds are predicted to form in larger numbers than heavy seeds, but light MBH seeds have more difficulty binding in binaries, and therefore their merger rate is suppressed. Unfortunately this is a very degenerate problem: seed formation, growth and dynamics all play equally important roles. Care will be needed when interpreting LISA data.

\section*{Acknowledgements}

This work was supported by the CNES for the space mission LISA.  The authors thank the anonymous referee for a constructive review.  MV thanks Helvi Witek for stressing the importance of the mass ratio distribution of merging MBHs and Michael Tremmel for thoughtful comments. MV and MC would like to acknowledge networking support by the COST Action GWverse CA16104. This work was partially supported by the Segal grant ANR-19-CE31- 0017 (www.secular-evolution.org) of the French Agence Nationale de la Recherche and by the ANR grant LYRICS (ANR- 16-CE31-001 1). HP is indebted to the Danish National Research Foundation (DNRF132) and the Hong Kong government (GRF grant HKU27305119) for support. RJ acknowledges support from the STFC [ST/R504786/1]. SKY acknowledges support from the Korean National Research Foundation (NRF-2020R1A2C3003769). MT is supported by the Deutsche Forschungsgemeinschaft (DFG, German Research Foundation) under Germany's Excellence Strategy EXC 2181/1 - 390900948 (the Heidelberg STRUCTURES Excellence Cluster). 
This work was granted access to the HPC resources of CINES under the allocations 2013047012, 2014047012, 2015047012, c2016047637, A0020407637 made by GENCI and KSC-2017-G2-0003 by KISTI, and as  a ``Grand  Challenge''  project  granted  by  GENCI  on  the  AMD-Rome  extension of the Joliot Curie supercomputer at TGCC, and under the the allocation 2019-A0070402192 made by GENCI.
This work has made use of the Horizon Cluster hosted by Institut d'Astrophysique de Paris. We thank Stephane Rouberol for running smoothly this cluster for us. 

\section*{Data Availability}
The data underlying this article were provided by the Horizon-AGN and NewHorizon collaborations by permission. Data will be shared on request to the corresponding author with permission of the Horizon-AGN and NewHorizon collaborations.

\bibliographystyle{mnras}
\bibliography{biblio_complete} 

\appendix

\section{Evolving black hole and black hole properties in estimates of the dynamical friction timescale}
\label{app:evolution}

One assumption made in our estimate of the dynamical friction timescale $t_{\rm df}$ is that MBH and galaxy properties remain fixed after the time of the 'numerical' merger. In this section, we discuss the impact of this assumption using examples from \nh.

After the MBHs are numerically merged at time $t_{\rm in}$, their individual mass evolution is no longer tracked by the simulation. However, the new, 'numerically-merged' MBH does continue to grow due to accretion. We record the values of the MBH individual masses  and of the total mass  of the pair at $t_{\rm in}$, and compute the  accretion rate on each MBH and on the numerically-merged MBH using the  Bondi-Hoyle-Lyttleton (BHL) accretion rate:

\begin{equation}
\dot{M}_{\rm acc} = \frac{G^2 M_{\rm in}^2 \ \rho_{\rm gas}}{(c_s^2+v_{\rm rel}^2)^{3/2}},
\label{eq:BHL}
\end{equation}
where $M_{\rm in}$ is the mass of each MBH or of the numerically-merged binary at $t_{\rm in}$, $c_s$ and  $\rho_{\rm gas}$  the sound speed and gas density in the vicinity of the numerically merged pair, and $v_{\rm rel}$ the relative velocity between gas and the MBH at $t_{\rm in}$.  In \nh~the spatial resolution is high enough that no boost is included. 

\begin{figure}
    \centering
    \includegraphics[width=\columnwidth]{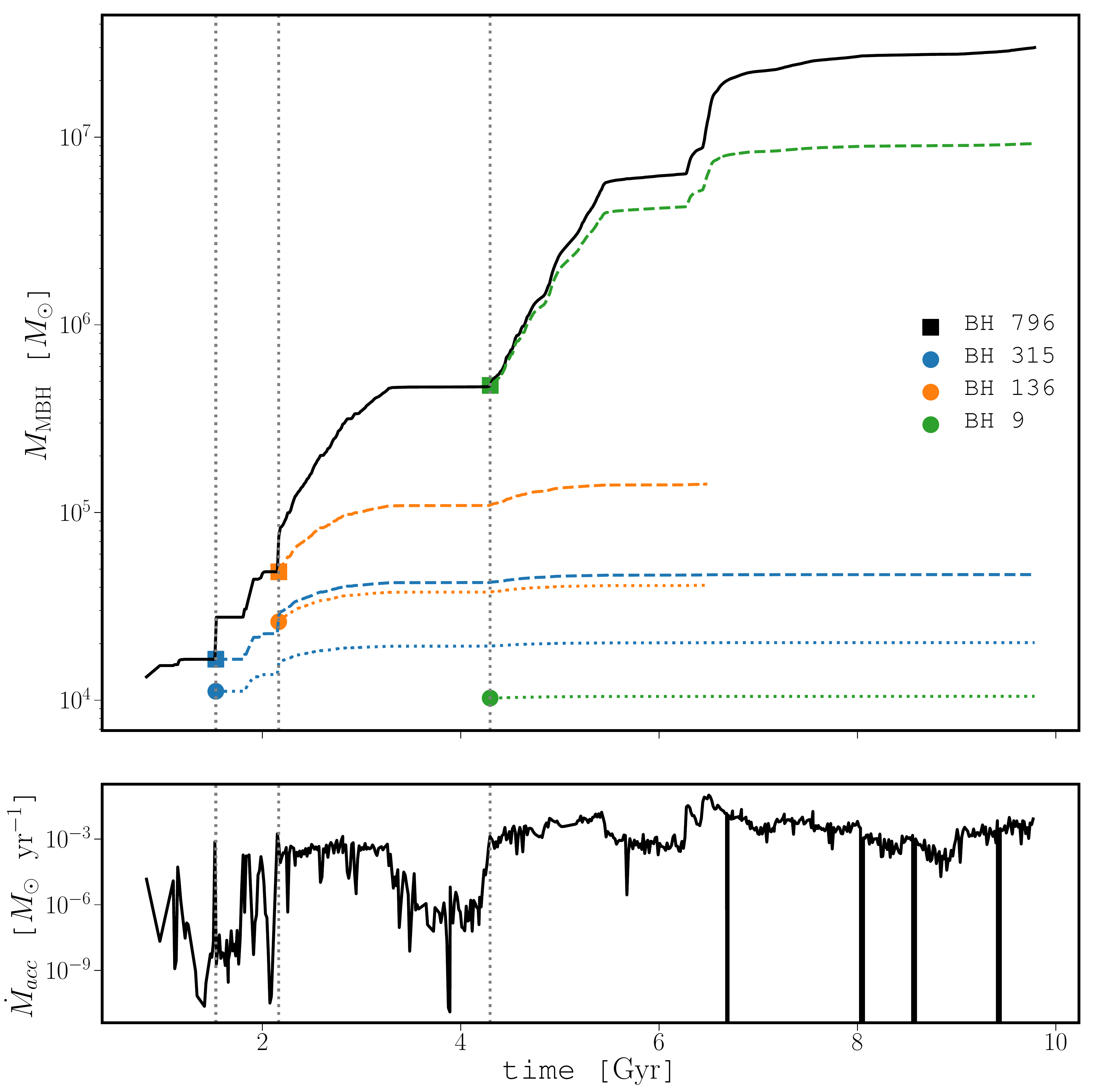}
    \caption{Top panel: Extrapolating MBH mass growth from the numerically merged MBH mass evolution (solid line), for primary (dashed lines) and secondary (dotted lines) MBHs during the first three mergers of MBH 796, for the timespan $t_{in}$ to $t_{in}+t_{df}$, or until the end of the simulation. Bottom panel: Accretion rate of MBH 796 over time. }
    \label{fig:mass_growth}
\end{figure}

In Figure \ref{fig:mass_growth} we show the post-processed mass evolution, using the first three mergers in which the primary is MBH 796 in \nh, as an example.
 Each mass evolution is integrated from the time of the numerical merger, $t_{\rm in}$ until $t_{\rm in}+t_{\rm df}$, with the primary (dashed) and secondary (dotted) black hole tracked separately. Also shown is the mass evolution of the numerically merged MBH (solid line), from which the accretion rate (bottom panel) is calculated.

\begin{figure}
    \centering
    \includegraphics[width=\columnwidth]{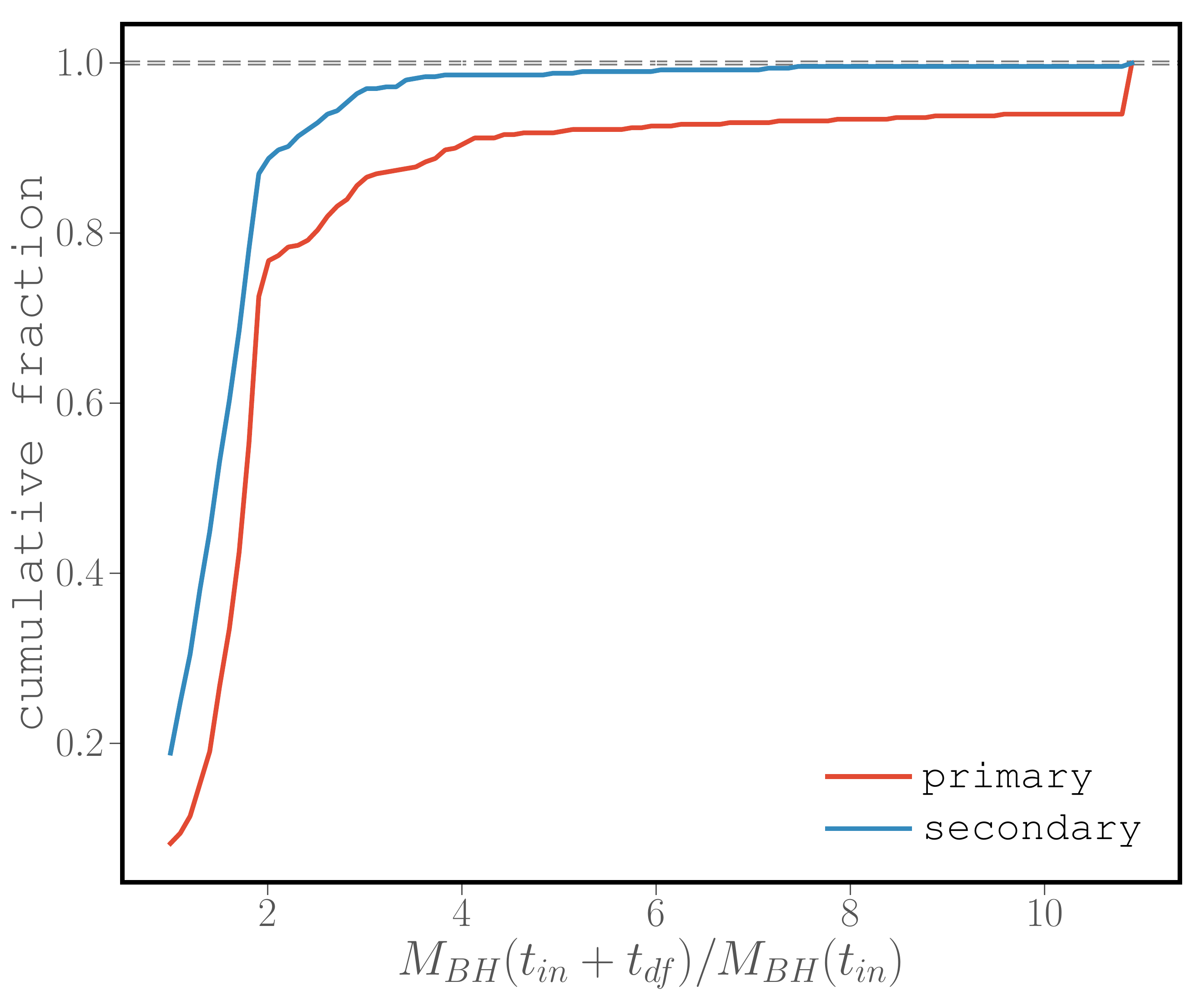}
    \caption{Cumulative distribution of mass increases during $t_{df}$, following all mergers in \nh.}
    \label{fig:cumulative_mass}
\end{figure}

As can be seen in Fig. \ref{fig:mass_growth}, the mass growth of both the primary (always 796, dashed lines) and the secondary MBHs (315, 136 and 9 respectively, dotted lines) from the point of numerical merger is small even over Gyr timescales, if one assumes that the individual black holes continue growing at the local Bondi-Hoyle-Littleton accretion rate. Even the primary grows significantly more slowly without the mass-boost provided during the numerical merger, due to the cumulative term of the $M^2_{BH}$ dependence of Eq. \ref{eq:BHL}.

Looking at the whole sample for \nh, Fig \ref{fig:cumulative_mass} shows that for the majority of the sample, mass growth is small during $t_{\rm df}$ for both the primary and the secondary. For \nh, 87\% (72.5\%) of secondary (primary) MBHs less than double their mass by $t_{\rm df}$, while 98.8\% (91.8\%) gain less than a factor of 5 in mass. As this extra mass will accumulate progressively, the total mass gain during $t_{\rm df}$ sets an upper limit on how much $t_{\rm df}$ would vary if we had taken continuous mass growth into account. We therefore conclude that the impact of calculating $t_{\rm df}$ using the black hole masses at the time of numerical merger is small overall.

Conclusions are similar when studying the impact of evolving  the galaxy stellar mass  $M_{\rm gal}$ and velocity dispersion $\sigma$. From Eq. \ref{eq:tdf}, the dynamical friction time $t_{\rm df}$ depends logarithmically on $M_{\rm gal}$ and thus very weakly on the stellar mass. The dependence is linear for $\sigma$, but the impact of growing $\sigma$ is more difficult to quantify.

If to first approximation we describe the host galaxy as an isothermal sphere evolving toward higher masses and thus higher stellar velocity dispersion preserving spherical symmetry, 
no torque is acting on the MBH (treated as test particle), despite the change in $\sigma$, leading to the conservation of its specific angular momentum ${\tilde L}= r\,v_c$.
Since the circular velocity $v_c = \sqrt{2} \sigma$ is independent of the enclosed mass and radius $r$, the orbiting MBH responds by reducing the radius $r$ to compensate the increase in $v_c$ required by the new dynamical equilibrium condition.  Since the frictional torque $\tau_{\rm df}$ on a test mass in a circular orbit  is equal to $\tau_{\rm df}=-0.428GM_{\rm BH}/r$, we can define a `reference' dynamical friction timescale $t^{0}_{\rm df}=1.17 r_0^2 v_{c,0}/(\ln \Lambda GM_{\rm BH})$ for a MBH at distance $r_0$ with circular velocity $v_{c,0}$ in an unperturbed isothermal sphere \citep{binney1987}. If, as an example, we assume that the circular velocity increases 
exponentially, over a timescale $t_{\rm gal},$ i.e., $v_c(t)=v_{c,0}\exp(t/t_{\rm gal})$, then the evolution equation for the radius ${\dot r}v_c+r{\dot v}_c =\tau_{\rm df}$ 
leads to orbital decay down to $r=0$ over a timescale
\begin{equation}
    t_{\rm df}=t_\mathrm{gal} \ln\left( 1+t_\mathrm{df}^\mathrm{0}/ t_\mathrm{ gal} \right),
    \label{expo}
\end{equation}
which is always smaller than $t^{0}_{\rm df}/t_\mathrm{gal}.$ 
Likewise, if we consider a power-law increase in $\sigma$ and thus in the circular velocity of the form $v_c=v_{c,0}(1+t/t_{\rm gal})^\beta$, orbital decay to $r=0$ occurs on a timescale
\beq
t_{\rm df}=t_\mathrm{gal}\left[ \left(1+(\beta+1)\frac{t^0_{\rm df}}{t_{\rm gal}}\right)^{1/(\beta+1)} -1 \right],
\eeq
which for $\beta>0$ implies again values of the dynamical friction time smaller than $t^{0}_{\rm df}/t_\mathrm{gal}.$

If  $t^0_{\rm df}$ is short compared to $t_{\rm gal}$  the evolution of the host galaxy properties during this time will be negligible, and $ t_{\rm df}\sim t_{\rm df}^0$.  Otherwise its value is determined by the timescale $t_{\rm gal}$ over which  $v_c$ and accordingly $\sigma$ increase. Thus, $t_{\rm df}^0$ gives an upper limit on the dynamical friction timescale, in this simplified toy model. Since, even assuming galaxies can be described as isothermal spheres, the mass growth of individual galaxies in time generally does not follow a simple analytical expression, we refrain from implementing an arbitrary general correction to the dynamical friction timescale. We therefore calculate all values of $t_{\rm df}$ using the MBH and galaxy properties at the time of numerical merger. Furthermore, if most of the galaxy growth is through the accretion of gas with large angular momentum, which forms extended discs, the properties of the central regions of the galaxies should be little affected, which also means that the dynamical friction timescales of the MBH should not greatly be affected even if the galaxy grows significantly.

\section{Stellar density and binary hardening}
\label{app:gamma}

\begin{figure}
    \includegraphics[width=\columnwidth]{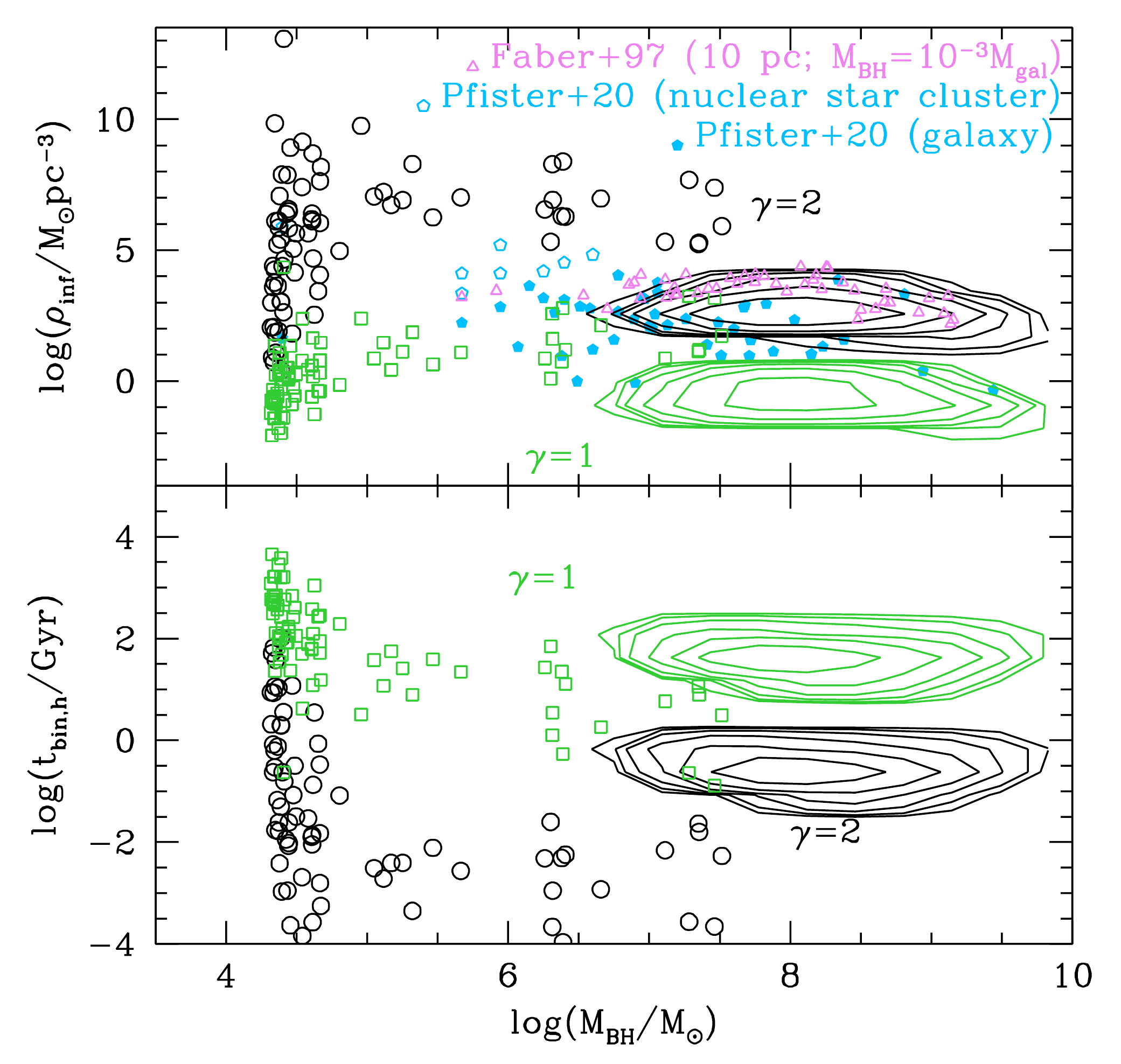}
    \caption{Stellar density at the MBH sphere of influence (top) and binary hardening timescales (bottom) for MBHs in \nh~(black circles for $\gamma=2$ and green squares for $\gamma=1$) and \hagn~(black and green contours for $\gamma=2$ and $\gamma=1$ respectively). The estimates for nearby galaxies are from \citet{Faber1997} and \citet{2020arXiv200308133P}, see text for details.}
    \label{fig:gamma}
\end{figure}

In the body of the paper we have shown results for binary evolution timescales when the central stellar density profile of galaxies is modelled as a power-law with $\gamma=2$ (isothermal sphere). We show in Fig.~\ref{fig:gamma} $t_{\rm bin,h}$ and $\rho_{\rm inf}$ for $\gamma=2$ and $\gamma=1$, recalling that:
\begin{equation}
r_{\rm inf}=R_{\rm eff}\left(\frac{4M_{BH}}{M_{\rm gal}}\right)^{1/(3-\gamma)}
\label{eq:rinf}
\end{equation}
and 
\begin{equation}
\rho_{\rm inf}=\frac{(3-\gamma)M_{\rm gal}r_{\rm inf}^{-\gamma}}{8\pi R_{\rm eff}^{3-\gamma}}.
\label{eq:rhoinf}
\end{equation}

We compare with $\rho_{\rm inf}$ estimated by \cite{2020arXiv200308133P} using a Prugniel profile \citep{Prugniel_97}, for a sample of 45 nearby galaxies with measured MBH mass and bulge properties (filled pentagons), some of them including also a nuclear star cluster (empty pentagons) as well as the density at 10~pc estimated by \cite{Faber1997} on a sample of 46 elliptical galaxies using a Nuker profile \citep[][]{1995AJ....110.2622L}; for these elliptical galaxies we assume that the MBH mass is $10^{-3}$ the mass of the galaxy (violet triangles). The inner 3D logarithmic slope is by construction $<1$ for a Prugniel profile while the deprojected Nuker profile allows for larger values. 

The densities estimated with $\gamma=2$ are in good agreement with observations for \hagn, while for \nh~they could be higher than the values of $z=0$ MBHs, although this is not completely clear: on the one hand the 10~pc radius used in \cite{Faber1997} is much larger than $r_{\rm inf}$ for \nh~MBHs, on the other hand the Prugniel profile used in \cite{2020arXiv200308133P} has by construction a shallow inner slope. In any case, the higher densities in \nh~compared to \hagn~is a result of the MBHs being lighter in \nh~than in \hagn~at fixed galaxy mass (see Equations~\ref{eq:rinf} and~\ref{eq:rhoinf}), and of galaxies in \nh~being more compact, i.e., having smaller $R_{\rm eff}$ at a given $M_{\rm gal}$.

\section{Removing spurious mergers}
\label{app:spurious}

To identify possible spurious numerical mergers we selected MBHs within $\max(2R_{\rm eff},4\Delta x)$ (``\rbg''), at each step, i.e., this condition was applied at the time of the numerical merger ($t_{\rm in}$), after the dynamical friction timescale ($t_{\rm in}+t_{\rm df}$), and after the binary evolution timescale ($t_{\rm in}+t_{\rm df}+t_{\rm bin}$).

\begin{figure*}
    \begin{overpic}[width=0.45\textwidth]{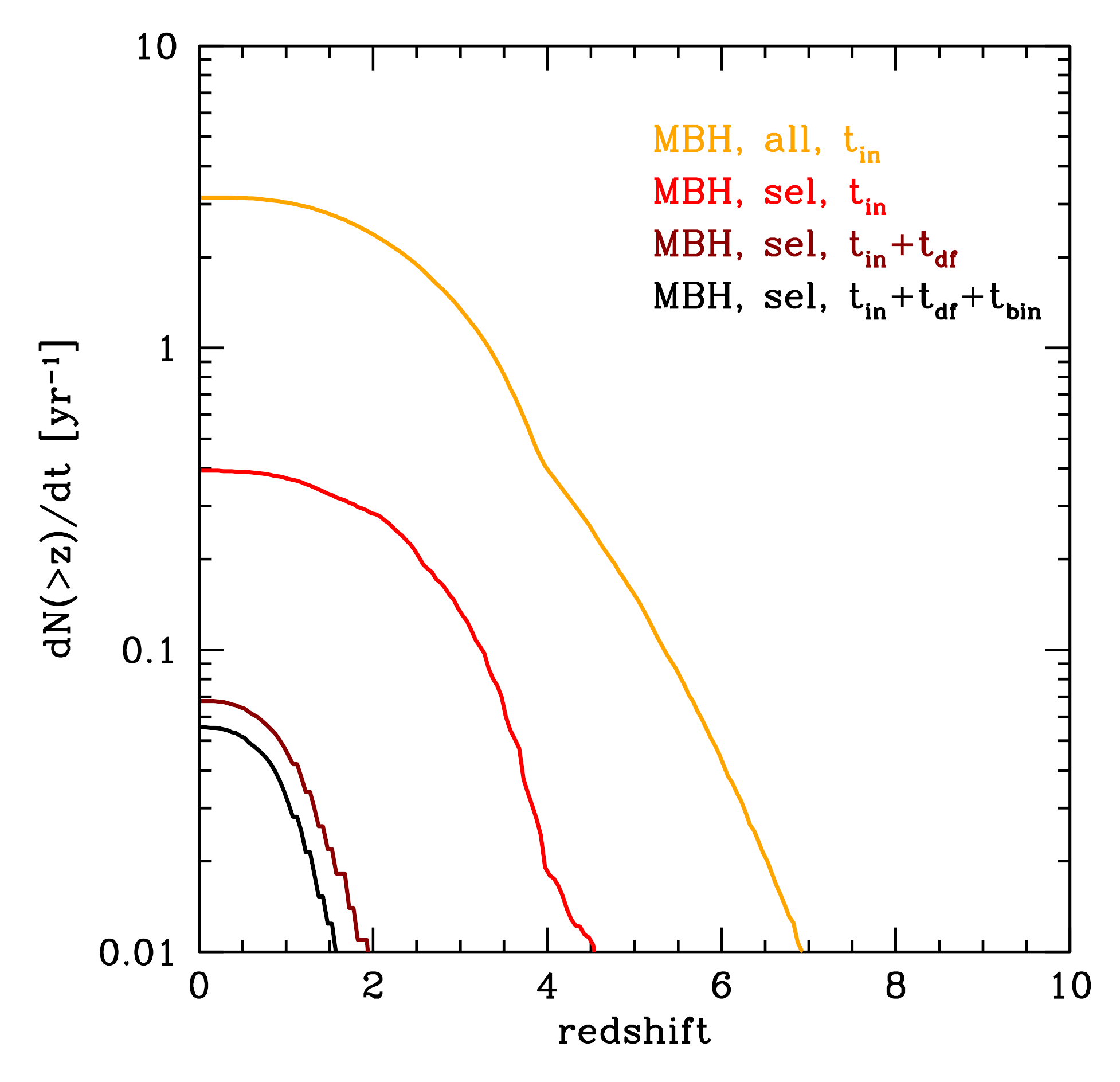} 
	\put(22,88){\hagn}
	\end{overpic}
 	\includegraphics[width=0.45\textwidth]{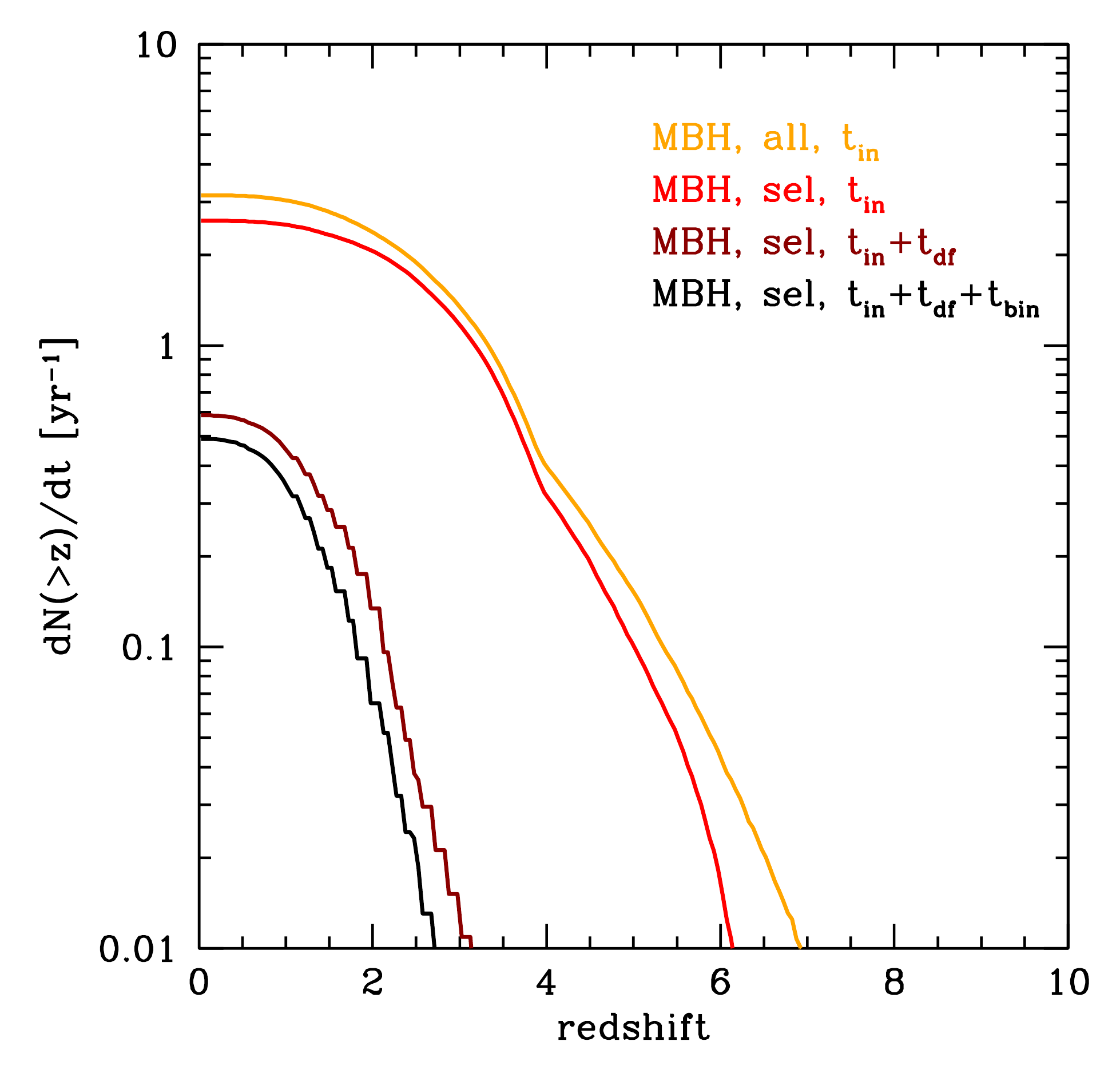}
    \hfill
    \begin{overpic}[width=0.45\textwidth]{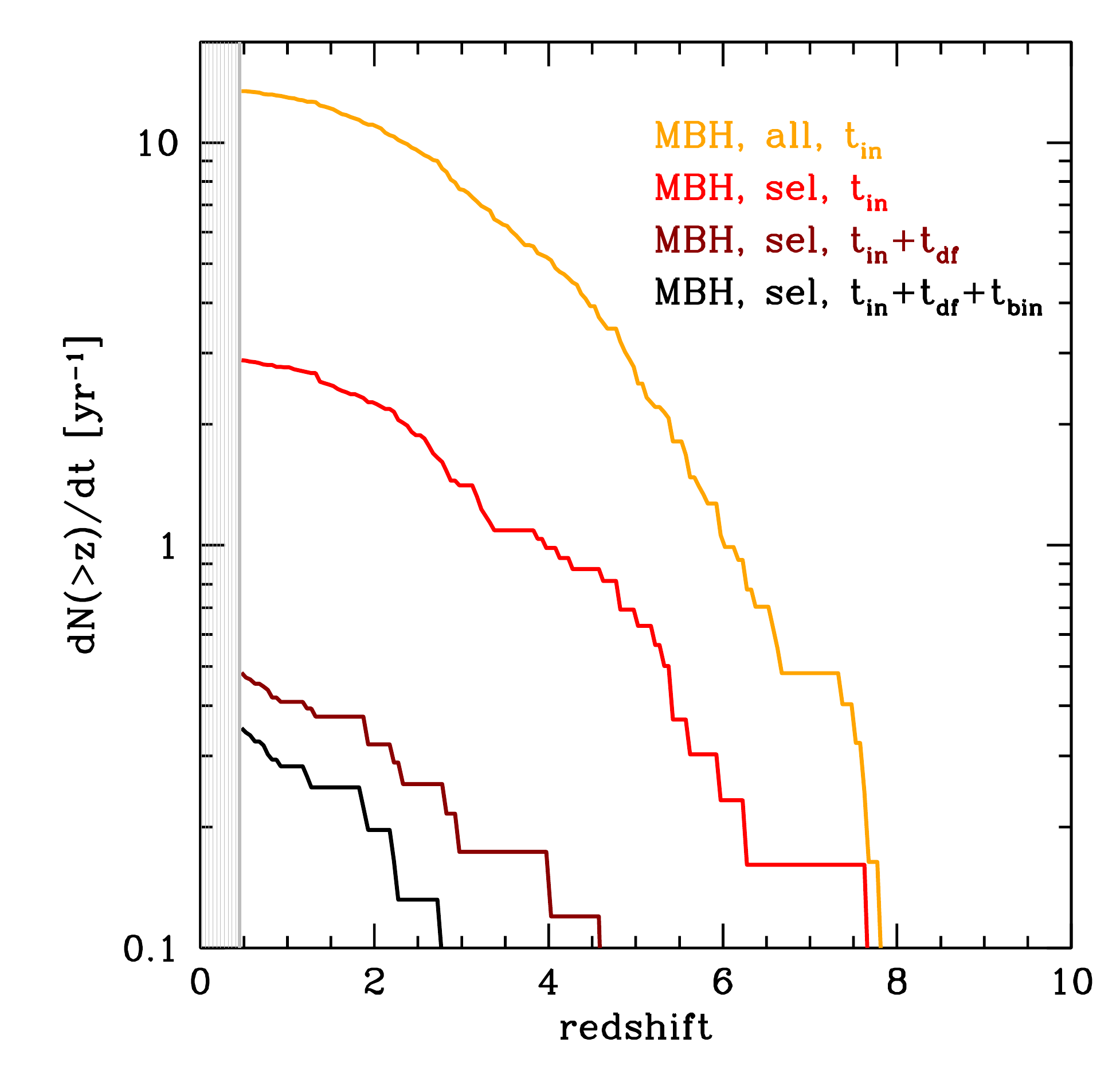}
	\put(22,88){\nh}
    \end{overpic}
 	\includegraphics[width=0.45\textwidth]{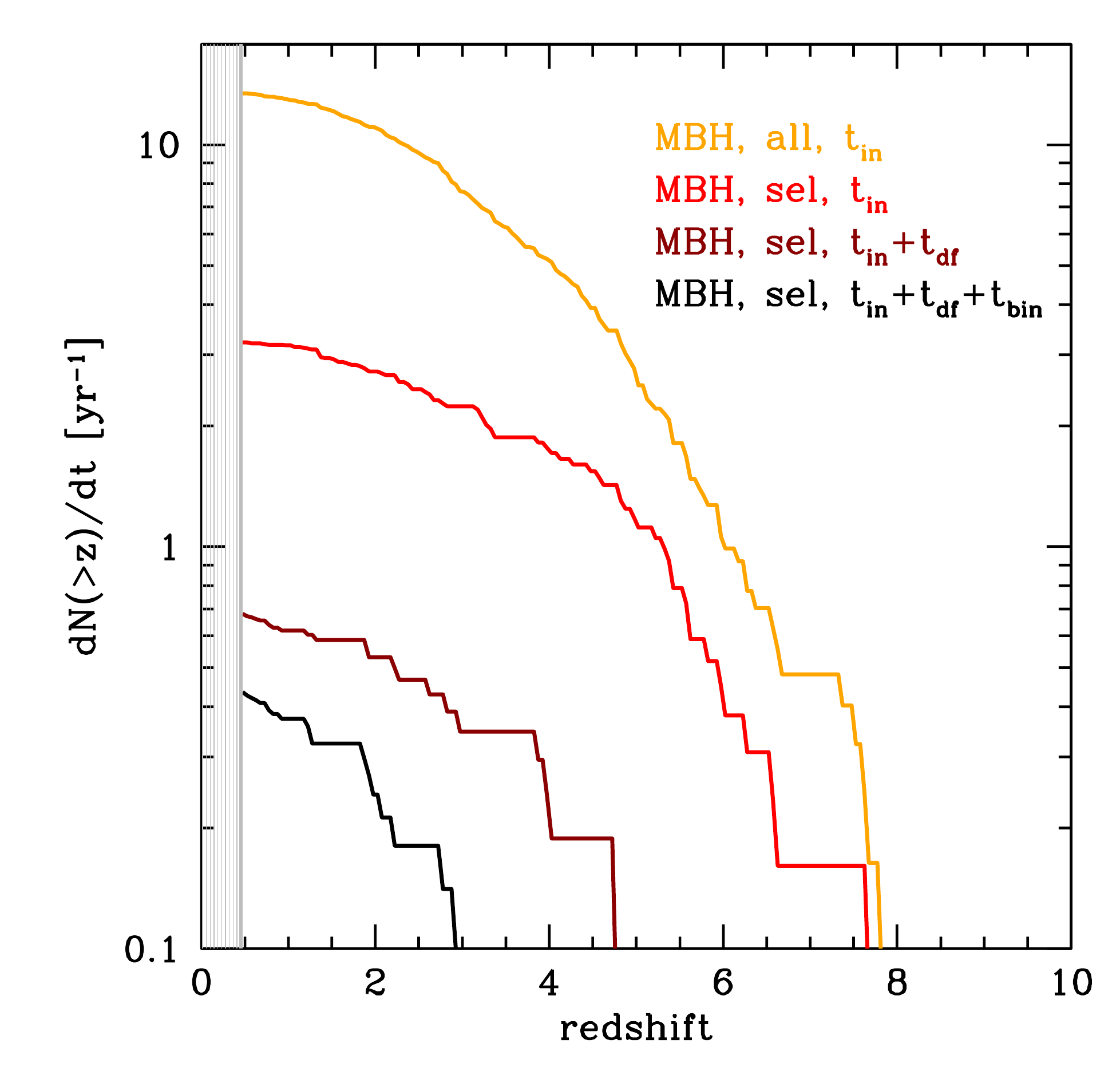}
\caption{Analogue of Fig.~\ref{fig:BH_mergrate} but for catalogues \rsm~and \dx, from left to right.}
    \label{fig:BH_mergrate_app}
\end{figure*}

We here discuss alternative choices for the criterion at the same times: (i) pairs where the MBHs are within $8 \Delta x$ from the centre of the nearest galaxy (``\dx'') at the time of the numerical merger, (ii) MBHs within $\max(0.5R_{\rm eff},4\Delta x)$ (``\rsm''). In \nh, these catalogues contain 43 and 81 MBHs, and in \hagn~18500 and 1329  MBHs. Note that catalogue \rsm~is a subset of \rbg, but catalogue \dx~is not necessarily a subset of any of the two: in high-redshift galaxies, $R_{\rm eff}$ can be smaller than $8 \Delta x$, especially in the case of \hagn, while the opposite is true for massive, large low-redshift galaxies. A numerical size, \dx, instead of a physical galaxy size can be useful at early cosmic times, when galaxies are very ``messy'' and defining the galaxy centre challenging. 

We show in Fig.~\ref{fig:BH_mergrate_app} the MBH merger rates resulting from these alternative catalogues. The merger rate for catalogue \dx~at later times converges towards that of \rbg~for \hagn~and \rsm~for \nh. The reason is that $8 \Delta x$ represents very different physical scales for the two simulations, 8~kpc and 320~pc, the former closer to the full extent of massive ($M_*>10^{11}\, \rm \msun$) low-redshift galaxies~\citep{Dubois2016} and the latter closer to the size of the central region, somewhat smaller than typical bulge sizes. 

The other results of the paper are not very different if we consider \rsm~instead of \rbg, except for the obvious smaller number of MBHs. We note only that the masses of merging galaxies sourcing MBH mergers in \rsm~are somewhat larger than in \rbg, 0.3 dex in \hagn~and 0.62 dex in \nh. MBHs are more centrally located in more massive galaxies, which have a smoother and deeper potential wells.

\bsp	
\label{lastpage}
\end{document}